\documentclass[iop,apj,twocolappendix,numberedappendix]{emulateapj}
\usepackage[varg]{txfonts}
\usepackage{bm}
 \usepackage{color}

\newcommand{\beq}{\begin{equation}}
\newcommand{\eeq}{\end{equation}}
\newcommand{\beqn}{\begin{eqnarray}}
\newcommand{\eeqn}{\end{eqnarray}}

\newcommand{\eqref}[1]{(\ref{#1})}
\newcommand{\eqn}[1]{Equation~(\ref{eq:#1})}

\newcommand{\dfrac}[2]{ {\displaystyle\frac{#1}{#2}} }

\newcommand{\pfrac}[2]{ \biggl(\dfrac{#1}{#2}\biggr) }

\newcommand{\pd}{\partial}
 
\newcommand{\kB}{k_{\rm B}}

\newcommand{\brac}[1]{\langle #1 \rangle}
\newcommand{\dbrac}[1]{\left\langle\left\langle #1 \right\rangle\right\rangle}
\newcommand{\eps}{\epsilon}

\newcommand{\Ecrit}{E_{\rm crit}}
\newcommand{\Ecriti}{E_{{\rm crit},i}}
\newcommand{\Bstar}{{${\rm B}^*$}}
\newcommand{\Cstar}{${\rm C}^*$}

\newcommand{\IP}{{\rm IP}}

\shorttitle{Nonlinear Ohm's Law}
\shortauthors{Okuzumi \& Inutsuka}

\begin{document}
\title{The Nonlinear Ohm's Law: Plasma Heating by Strong Electric Fields and its Effects on the Ionization Balance in Protoplanetary Disks}
\email{}
\author{Satoshi Okuzumi\altaffilmark{1,2} and Shu-ichiro Inutsuka\altaffilmark{2}}
\affil{$^1$Department of Earth and Planetary Sciences, 
Tokyo Institute of Technology, Meguro-ku, Tokyo, 152-8551, Japan; okuzumi@geo.titech.ac.jp\\
$^2$Department of Physics, Nagoya University, Nagoya, Aichi 464-8602, Japan}

\begin{abstract}
The ionization state of the gas plays a key role in the MHD of protoplanetary disks. However, the ionization state can depend on the gas dynamics, because electric fields induced by MHD turbulence can heat up plasmas and thereby affect the ionization balance. To study this nonlinear feedback, we construct an ionization model that includes plasma heating by electric fields and impact ionization by heated electrons, as well as charging of dust grains. We show that when plasma sticking onto grains is the dominant recombination process, the electron abundance in the gas decreases with increasing electric field strength. This is a natural consequence of electron-grain collisions whose frequency increases with electron's random velocity. The decreasing electron abundance may lead to a self-regulation of MHD turbulence. In some cases, {not only the electron abundance but also the electric current decreases} with increasing field strength in a certain field range. The {resulting} N-shaped current-field relation violates the fundamental assumption of the non-relativistic MHD that the electric field is uniquely determined by the current density. {At even higher field strengths,} impact ionization causes an abrupt increase of the electric current as expected by previous studies. We find that this discharge current is multi-valued (i.e., the current--field relation is S-shaped) under some circumstances, and that the intermediate branch is unstable. {The N/S-shaped current--field relations may yield hysteresis in the evolution of MHD turbulence} in some parts of protoplanetary disks.
\end{abstract}
\maketitle

\section{Introduction}
How protoplanetary disks form and evolve is a key question of planet formation studies. 
It is generally accepted that magnetic fields play many important roles in these processes.
Coupling between a disk and a magnetic field induces 
magnetorotational instability \citep[MRI;][]{BH91}, 
and turbulence driven by MRI provides a high effective viscosity that allows disk accretion \citep[e.g.,][]{HGB95,FN06,FDK+11}. 
{A large-scale magnetic field threading a disk also drives outflows from disk surfaces 
via the recurrent breakup of MRI modes \citep[][]{SI09,SI14} and/or magnetocentrifugal mechanism 
\citep[e.g.,][]{BP82,S96,LFO13,FLLO13,BS13a}. 
These outflows can significantly affect disk structure and planet formation in inner disk regions \citep{SMI10}.
}

In protoplanetary disks, 
these magnetic activities are strongly subject to non-ideal MHD effects 
simply because  the  ionization fraction of the disk gas is low.
Thermal ionization is effective only in innermost disk regions where 
the gas temperature exceeds 1000 K \citep{U83}.
Further out, the disk gas is only weakly ionized by external ionizing sources 
including cosmic rays \citep{UN81} and stellar X-rays \citep{GNI97}.
The resulting high ohmic conductivity yields an MRI stable ``dead zone'' 
near the midplane \citep{G96,SMUN00}.
Ambipolar diffusion also suppresses MRI near the disk surface \citep{PC11a,B11a,SBA+13a,SBA+13b},
and the combined effect of ohmic and ambipolar diffusion can render MRI inactive at all altitudes
in inner disk regions \citep{BS13b,B13,LKF14}. 
The high diffusivities also cause the loss of large-scale magnetic fields that are 
required for magnetocentrifugal outflow \citep[e.g.,][]{LPP94a}.
Hall drift can either enhance or suppress the magnetic activity of the disks depending 
on the polarity of the large-scale magnetic field relative to disk's rotation axis \citep{WS12,LKF14,B14}.

At the same time, these magnetic activities can influence the ionization state of the disk gas.
MRI turbulence transports ionized gas to less ionized regions,
and this process revives MRI in dead zones under favorable conditions \citep{IS05,TSD07,IN08}. 
Joule heating by MRI turbulence can change the temperature profile of the disks
and even the location of the dead-zone inner edge \citep{LB12,FFL14}. 
On smaller scales, strong current sheets produced by MRI can locally heat up the disk gas,
which can even affect the ionization state if the background temperature is near the thermal ionization threshold 
\citep{HMM12,MHM+13,MHYM14}.  

This study focuses on the role of strong electric fields in the ionization balance.
Since the electric conductivity of the disk gas is finite, the coupling between 
the moving gas and magnetic fields {inevitably} produces 
a nonzero electric field in the comoving frame of the gas.
The field induces systematic drift motions of ionized gas particles,
whose effect is expressed by the conventional Ohm's law in which 
the electric current is linearly proportional to the electric field strength. 
However, what is largely unappreciated is that in an weakly ionized plasma, 
an electric field also induces random motion of plasma particles 
if the field is sufficiently strong \citep{DP40,GZS80,LP81}.  
In principle, the heating of plasmas affects the chemical reactions of the plasmas,
thereby affecting the ionization balance of the gas. 
This effect has been ignored by all previous models of disk ionization.

\citet{IS05} first pointed out that this electric plasma heating can occur  
in protoplanetary disks when MRI drives disk turbulence.
They noted that MRI-driven turbulence produces a strong electric field
in the neutral comoving frame when the ionization degree is low 
(but not too low for MRI to be active). 
They estimated the random energy of electrically heated free electrons, 
and concluded that the energy can be high enough to cause electric discharge 
(or electron avalanche) in the disk gas.
They suggested that MRI in protoplanetary disks can be self-sustained: 
MRI turbulence can provide sufficient ionization to keep MRI active
even in the conventional dead zones.
This scenario has been recently tested by \citet{MOI12} using local MHD simulations
with a toy resistivity model that mimics electron discharge at high electric field strengths.
They found that self-sustained MRI is realized when the work done 
by the turbulence exceeds the energy consumed by Joule heating.

In this paper, we study in detail the effects of plasma heating by electric fields 
on the ionization balance of a weakly ionized gas.  
While the discharge is only produced by very high electric fields,  
weaker fields still can heat up plasmas and can change the reaction balance.
To reveal the consequences of electric plasma heating over a wide range 
of electric field strengths, we construct a charge reaction model 
properly taking into account the kinetics of weakly ionized plasmas
 under an electric field \citep{DP40,GZS80,LP81}.
Our model also includes plasma capture by small dust grains, which 
is essential to study the ionization balance in dense protoplanetary disks 
\citep[e.g.,][]{U83,SMUN00,IN06a,W07,B11a}.
As a first step, we neglect the effects of magnetic fields on the kinetics of plasmas,
which means that we only treat ohmic conductivity and neglect ambipolar diffusion and Hall drift. 
An extension of our model to the non-ohmic resistivities will be done in future work.

The paper is organized as follows.
In Section~\ref{sec:estimate}, we present an order-of-magnitude estimate 
of the electric field strength in MRI-driven turbulence to highlight the potential 
importance of electric plasma heating in weakly ionized protoplanetary disks.
Our charge reaction model is described in Section~\ref{sec:model}, 
and results are presented in Sections~\ref{sec:nodischarge} and \ref{sec:discharge}.
Section~\ref{sec:discussion} discusses important implications 
for MHD in protoplanetary disks. 
A summary is given in Section~\ref{sec:summary}.

\section{Plasma Heating by MRI Turbulence in Protoplanetary Disks}\label{sec:estimate}
Before presenting our charge reaction model, 
we briefly describe the basic physics of electric plasma heating in a weakly ionized gas.
We will then demonstrate by simple estimations that the plasma heating can occur 
in protoplanetary disks under realistic conditions. 

Let us consider a weakly ionized gas in which neutral gas particles 
are much more abundant than plasma particles.
In such a gas, plasma particles collide with neutral particles 
much more frequently than with themselves. 
Therefore, if there is no externally applied field, 
the plasma particles tend to be thermally equilibrated with the neutrals,
and their mean kinetic energy approaches that of the neutrals, $3\kB T/2$, 
where $T$ is the neutral temperature and  $\kB$ is the Boltzmann constant.
However, if there is an applied electric field, 
the field accelerates the plasma particles, and some part of the gained energy 
is converted to their random energy after collisions with neutrals. 
This electric heating is particularly significant for electrons because of their high mobility and 
low energy transfer efficiency in collisions with neutrals.
In equilibrium, the random energy of electrons greatly exceeds that of neutrals ($3\kB T/2$) 
when $E$ is well above the threshold 
\beq
\Ecrit \equiv \sqrt{\frac{6m_e}{m_n}}\frac{\kB T}{e\ell_e},
\label{eq:Ecrit}
\eeq
where $e$ is the elementary charge and 
$m_e$ and $m_n$ are the masses of an electron and a neutral, respectively \citep{DP40,GZS80,LP81}.  
The electron mean free {path} $\ell_e$ is determined by the collisions with neutrals and is 
given by $\ell_e = 1/(n_n\sigma_{en})$, where $n_n$ is the neutral number density and
$\sigma_{en}$ is the momentum-transfer cross section for electron--neutral collisions.
{Equation~\eqref{eq:Ecrit} neglects energy losses due to inelastic electron--neutral collisions 
(i.e., collisions that involve electronic/vibrational/rotational excitation of the neutrals)
and radiative energy losses upon collisions with positive ions. 
The former is negligible at least at the onset of electron heating (for details, see Section~\ref{sec:velocity})
and the latter is generally negligible in weakly ionized plasmas where electron--ion collision are rare.} 
The small factor $\sqrt{m_e/m_n}$ {in Equation~\eqref{eq:Ecrit}} comes from the fact that 
electrons lose only a small fraction ($\sim m_e/m_n$) of their kinetic energy 
in a single {\it elastic} collision a neutral (for details, see Appendix~\ref{sec:kinetics}).  
For an ${\rm H}_2$ gas, $\sigma_{en} \approx 10^{-15}~{\rm cm^2}$ 
at electron energies $< 10~{\rm eV}$ \citep{FP62,YSH+08}, so we have
\beq
E_{\rm crit} \approx 1\times 10^{-9} \pfrac{T}{100~{\rm K}} \pfrac{n_n}{10^{12}~{\rm cm^{-3}}}~{\rm esu~cm^{-2}}.
\eeq

Let us see whether MRI turbulence in protoplanetary disks can provide such a strong electric field.
We denote the mean amplitude of the electric field in MRI turbulence, as measured in the comoving 
frame of the neutral gas, by $E_{\rm MRI}$.\footnote{
Throughout this paper, we refer to electric fields as measured in the comoving frame of the neutral gas.  
The electric field ${\bm E}$ in the comoving frame is related to the field ${\bm E}_{\bm u}$
in the frame where the gas moves at velocity ${\bm u}$ as 
${\bm E}_{\bm u} = {\bm E} - {\bm u}\times{\bm B}/c$, where ${\bm B}$ is the magnetic field.
}
To evaluate $E_{\rm MRI}$, we use the finding by \citet{MOI12} 
that the mean current density in MRI turbulence 
is insensitive to the strength of the ohmic resistivity. 
They performed local unstratified resistive MHD simulation and 
found that the current density in MRI turbulence has a mean amplitude
\beq
J_{\rm MRI} \approx f_{\rm sat}J_{\rm eqp}, 
\label{eq:JMRI}
\eeq
where $f_{\rm sat} \approx 10$ is a numerical constant and
\beq
J_{\rm eqp} \equiv \sqrt{\frac{\rho_g}{2\pi}} c\Omega
\label{eq:Jeqp}
\eeq
depends only on the gas mass density $\rho_g = m_nn_n$ and orbital frequency $\Omega$.\footnote{Equation~\eqref{eq:JMRI} can also be derived from 
an order-of-magnitude estimate of Ampere's law, ${\bm J} = (c/4\pi)\nabla\times {\bm B}$. 
Let us assume that the magnetic field associated with MRI has 
the characteristic wavenumber $k_{\rm MRI}$ and mean amplitude $B_{\rm MRI}$. 
Then, an order-of-magnitude estimate of Ampere's law gives 
$J_{\rm MRI} \sim (c/4\pi)k_{\rm MRI} B_{\rm MRI}$. 
For MRI-driven turbulence, $k_{\rm MRI}$ is comparable to that of the most unstable MRI modes,  
$k_{\rm MRI} \sim \Omega/v_{Az}$.
Since $v_{Az} \sim B_{z,\rm MRI}/\sqrt{4\pi \rho_g}$, where 
$B_{z,\rm MRI}$ is the vertical component of the fluctuating magnetic field, 
we have $J_{\rm MRI} \sim (B_{\rm MRI}/\sqrt{2}B_{z,\rm MRI})J_{\rm eqp}$. 
This reduces to Equation~\eqref{eq:JMRI} if $B_{\rm MRI}/\sqrt{2}B_{z,\rm MRI} \sim f_{\rm sat}$.
}
The value of $f_{\rm sat}$ is independent of the strength of the ohmic resistivity 
as long as sustained MRI turbulence is realized.\footnote{
We note, however, that $f_{\rm sat}$ can fall below $10$ 
when ambipolar diffusion is effective  \citep[see Figure 6 of][]{BS11}.
} The criterion for sustained turbulence can be given in terms of 
the Elsasser number 
\beq
\Lambda \equiv \frac{v_{Az}^2}{\eta\Omega},
\label{eq:Lambda}
\eeq
where $\eta$ is the ohmic resistivity and $v_{Az}$ 
is the Alfv\'{e}n speed in the direction perpendicular to the disk's midplane.
MRI grows when $\Lambda > \Lambda_{\rm crit}$,
where $\Lambda_{\rm crit} \approx 0.1$--$1$ \citep[e.g.,][]{SIM98,TSD07,MOI12}. 

Given $J_{\rm MRI}$, one can estimate $E_{\rm MRI}$ by using Ohm's law $E = (4\pi\eta/c^2)J$.  
Here it is useful to rewrite the ohmic diffusivity as 
$\eta = v_{Az}^2/\Lambda\Omega = 2c_s^2/\beta_z\Lambda\Omega$, 
where 
$\beta_z = 2c_s^2/v_{Az}^2$ is the plasma beta of the vertical magnetic field 
and $c_s = \sqrt{\kB T/m_n}$ is the sound speed.
If we use this expression, Ohm's law can be rewritten as 
\beq
E = \frac{8\pi}{\beta_z\Lambda\Omega}\pfrac{c_s}{c}^2 J.
\label{eq:ohm}
\eeq
This expression is useful because $\beta_z \sim 100$--$1000$ 
for fully saturated MRI turbulence \citep[e.g.,][]{HGB95,MS00,FP06,SI09}.  
From Equation~\eqref{eq:ohm}, 
we obtain 
\beqn
E_{\rm MRI} &=& \frac{8\pi}{\beta_z\Lambda\Omega}\pfrac{c_s}{c}^2 J_{\rm MRI}
\nonumber \\
&\approx& \dfrac{2\times 10^{-7}}{\Lambda}\pfrac{100}{\beta_z}\pfrac{f_{\rm sat}}{10} 
\pfrac{T}{100~{\rm K}} \pfrac{n_n}{10^{12}~{\rm cm^{-3}}}^{1/2}~{\rm esu~cm^{-2}}.~~
\eeqn

Thus, the ratio of $E_{\rm MRI}$ to $\Ecrit$ is
\beq
\frac{E_{\rm MRI}}{\Ecrit} 
\approx
\frac{200}{\Lambda}\pfrac{100}{\beta_z}\pfrac{f_{\rm sat}}{10}\pfrac{n_n}{10^{12}~{\rm cm^{-3}}}^{-1/2}.
\label{eq:Eratio}
\eeq
Note that the ratio is independent of the gas temperature $T$. 
We find that $E_{\rm MRI}$ exceeds $\Ecrit$
if 
\beq
\Lambda \la 200\pfrac{100}{\beta_z}\pfrac{f_{\rm sat}}{10}\pfrac{n_n}{10^{12}~{\rm cm^{-3}}}^{-1/2}.
\eeq
At the same time, $\Lambda \ga \Lambda_{\rm crit}$ is required for MRI turbulence to be sustained.
Combining these two criteria, we arrive at the condition for plasma heating in MRI turbulence,
\beq
\Lambda_{\rm crit} \la \Lambda \la 200\pfrac{100}{\beta_z}\pfrac{f_{\rm sat}}{10}\pfrac{n_n}{10^{12}~{\rm cm^{-3}}}^{-1/2}.
\eeq
Since $\Lambda_{\rm crit} \approx 0.1$--1, 
both the criteria are satisfied when $n_n \la 10^{14}$--$10^{18}~{\rm cm^{-3}}$
for $f_{\rm sat} \approx 10$ and $\beta_z \sim 100$--$1000$.

For protoplanetary disks,  the neutral gas density at the midplane is generally supposed to be 
$10^{9}$--$10^{15}~{\rm cm^{-3}}$.
Therefore, in the presence of MRI-driven turbulence, 
significant plasma heating can occur in some parts of protoplanetary disks.

\section{Ionization Model}\label{sec:model}
\begin{figure*}[t]
\epsscale{1.0}
\plotone{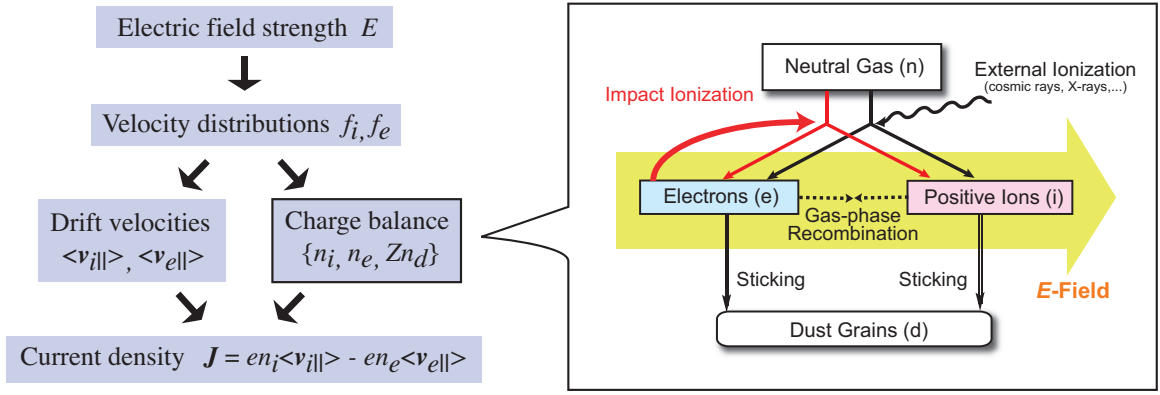}
\caption{Sketch of the charge reaction model presented in this study.
{The model gives the electric current density $J$ as a function of the electric field strength $E$.  
The plasma velocity distribution functions $f_i$ and $f_e$ used here take into account 
plasma heating by the electric field, and therefore the resulting $J$ is nonlinear in $E$. }
}
\label{fig:reaction}
\end{figure*}

In this section, we introduce a charge reaction model that takes 
into account heating of plasmas by strong electric fields. 
We consider a partially ionized dusty gas consisting of neutrals, singly charged positive ions, 
electrons, and dust grains. 
The gas is assumed to be so weakly ionized that the neutral number density $n_n$ 
can be approximated as a constant.
The neutral gas temperature $T$ is also assumed to be a constant by neglecting 
the Joule heating of the neutral gas (see \citet{MHYM14} 
for the possibility of significant local heating and thermal ionization of the neutral gas 
in MRI current sheets). 
We simplify the reaction network by representing the positive ions 
by a single dominant species (denoted by $i$).  
This one-component approach allows us to compute 
the ionization fraction of the gas without going into the details of the chemical composition of the ions,
and also with a good accuracy in particular 
when dust dominates the recombination process \citep{O09,I12}.
We will assume $i = {\rm HCO}^+$ in this study.  
Dust grains are spheres of single radius $a$ 
and are allowed to charge up by capturing plasma particles.
For simplicity, we do not consider size distribution of the grains in this study, 
but it is straightforward to do so because
the charge reaction only depends on a handful of moments of the size distribution \citep[see][]{O09}.

Charge reactions in the gas--dust mixture depend on the kinetic states of ions and electrons,
which are described by the velocity distribution functions $f_i$ and $f_e$, respectively. 
Unlike previous resistivity models, we give the distribution functions as a function of 
the electric field strength $E$ in the neutral rest frame.
We assume a steady state where acceleration by the electric field balances with 
energy/momentum losses upon collision with neutrals.
This assumption is valid for protoplanetary disks because 
collisions with neutral gas particles are much more frequent than 
charge reactions (which are collisions between plasma particles themselves or with dust grains) 
and also than the evolution of the electric field (which occurs on disk's dynamical timescale). 
The velocity distribution functions will be  presented in Section~\ref{sec:velocity}.

The charge reactions we consider are ionization by external high-energy particles 
(e.g., cosmic rays and X-rays), impact ionization by heated electrons, 
recombination of plasma particles in the gas phase, and plasma capture by dust grains.
The latter three reactions depend on the velocity distributions of the plasmas and hence 
on the electric field strength $E$.
Inclusion of impact ionization is essential to study electric discharge at high field strengths.
The rate equations and rate coefficients for the charge reactions will be given in Section~\ref{sec:reaction}.

One goal of this study is to reveal how the conventional Ohm's law is modified 
by strong electric fields. This can be done by calculating the current density as a function of 
the electric field strength $E$, and we do this in the following way 
(see Figure~\ref{fig:reaction} and also Section~\ref{sec:reaction}). 
First, calculate the mean drift velocities of plasma particles 
(denoted by $\brac{{\bm v}_{i||}}$ and $\brac{{\bm v}_{e||}}$) and the charge reaction rates as a function of $E$. 
We then calculate the ionization balance and obtain the number densities $n_i$ and $n_e$ 
of ions and electrons and the charge $Z$ of dust grains in equilibrium.
Finally, we obtain the current density as
\beq
{\bm J} ={\bm J}_i + {\bm J}_e,
\eeq
\beq
{\bm J}_\alpha \equiv q_\alpha n_\alpha\brac{{\bm v}_{\alpha||}} \quad (\alpha =i,e),
\eeq
where $q_i = e$ and $q_e = -e$ are the charges of ions or electrons, respectively.
We have neglected the contribution of charged grains to the ohmic conductivity  
because it is usually small \citep[see, e.g.,][]{B11b}.
Note that the resulting Ohm's law is nonlinear in $E$ because 
both $n_\alpha$ and $\brac{{\bm v}_{\alpha||}}$ depend on $E$,

\subsection{Velocity Distribution Functions and Their Moments}\label{sec:velocity}
Here we describe the velocity distribution functions of plasma particles  
and some averaged quantities (or ``velocity moments'') that will be used in later steps.   
We denote the velocity distribution functions for ions and electrons 
as $f_{i}({\bm E},{\bm v}_i)$ and $f_{e}({\bm E},{\bm v}_e)$, respectively, 
where ${\bm v}_\alpha$ $(\alpha = i,e)$ is the velocity of each ionized particle. 
{The first- and second-order moments of the distribution functions
give the mean drift velocity parallel to the electric field, $\brac{{\bm v}_{\alpha||}}$,
and the mean kinetic energy, $\brac{\eps_\alpha}$, as}
\beq
\brac{{\bm v}_{\alpha||}} 
= \hat{\bm E} \int (\hat{\bm E}\cdot{\bm v}_\alpha ) 
f_\alpha({\bm E},{\bm v}_\alpha) d^3v_\alpha
\label{eq:valpha||_def}
\eeq
{and}
\beq
\brac{\eps_\alpha} 
= \int \eps_\alpha f_\alpha({\bm E},{\bm v}_\alpha) d^3v_\alpha,
\label{eq:ealpha_def}
\eeq
respectively, 
where $\hat{\bm E} = {\bm E}/E$ and 
$\eps_\alpha = m_\alpha v_\alpha^2/2$  ($v_\alpha = |{\bm v}_\alpha|$).
{Note that the drift velocity perpendicular to ${\bm E}$ is zero in the absence of magnetic fields 
\citep[see, e.g.,][]{NU86a,W99}.}

As mentioned earlier, we assume that acceleration by electric fields is balanced with 
energy/momentum losses upon collisions with neutrals. 
In principle, a collision with a neutral is either ``elastic'' or ``inelastic,'' 
depending on whether the kinetic energy in the center-of-mass frame of the colliding particles 
is conserved or not (note, however, that both types of collisions can lead to energy loss of the 
charged particle in the neutral rest frame).
Inelastic energy losses are due to impact excitation (rotational/vibrational/electronic) and ionization of the neutrals.
However, these inelastic losses only enhance  
the efficiency of energy transfer from plasmas to neutrals by a factor of $\la 10$
(which is equivalent to increasing the electron mass by the same factor; see Appendix~\ref{sec:kinetics})
as long as the collision energy is $\la 1 {\rm eV}$ \cite[e.g., see Figure~15 of][]{EP63}.
This effect is particularly negligible at the onset of plasma heating (i.e., $E\sim \Ecrit$)
where the collision energy is $\sim \kB T \sim 10^{-2}~{\rm eV}$. 
For this reason, we neglect all inelastic losses and only consider elastic collisions 
in determining the velocity distributions of plasmas. 
This assumption allows us to use analytic expressions for the velocity distribution functions, 
which we will introduce below.

\subsubsection{Electrons}\label{sec:electron}
\begin{figure}[t]
\epsscale{1.1}
\plotone{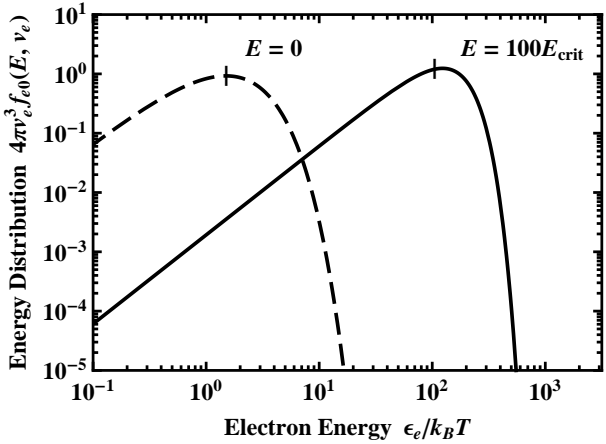}
\caption{Electron energy distribution 
$4\pi v_e^3 f_{e0}(E,v_e)$ as a function of the electron energy $\eps_e$ 
for $E = 0$ (dashed curve) and $E = 100\Ecrit$ (solid curve). 
The vertical ticks on the distributions indicate $\eps_e = \brac{\eps_e}$, 
where $\brac{\eps_e}$ is the mean electron energy (Equation~\eqref{eq:ee_exact}).
}
\label{fig:DF}
\end{figure}

Having neglected inelastic energy losses, one can analytically obtain 
the velocity distribution function for electrons in a weakly ionized gas 
using the Fokker-Planck (diffusion) approximation \citep[see][]{GZS80,LP81}. 
In the steady state, the distribution function is given by \citep{D35}
\beq
f_e({\bm E},{\bm v}_e) 
= \left(1-\frac{eE\ell_e}{\kB T}\frac{\eps_e \hat{\bm E}\cdot\hat{\bm v}_e}
{\eps_e+\chi \kB T}\right)f_{e0}(E,v_e),
\label{eq:fe}
\eeq
\beq
\chi \equiv \pfrac{E}{\Ecrit}^2 
\eeq 
where $\hat{\bm v}_e = {\bm v}_e/v_e$ and 
$f_{e0}$ is the ``symmetric'' part of $f_e$ 
{that depends on the magnitudes of ${\bm E}$ and ${\bm v}_e$ but not on 
the angle between them ($\cos^{-1}(\hat{\bm E}\cdot\hat{\bm v}_e)$)}.
The exact expression of $f_{e0}$ is 
\beq
f_{e0} = \pfrac{m_e}{2\pi \kB T}^{3/2}
\frac{({\eps_e}/{\kB T}+\chi)^\chi}{W(\chi)}\exp\left(-\frac{\eps_e}{\kB T}\right),
\label{eq:fe0}
\eeq
\beq
W(\chi) \equiv \chi^{3/2+\chi}U(\case{3}{2},\case{5}{2}+\chi,\chi),
\eeq
where 
$U(x, y, z) \equiv \Gamma(x)^{-1}\int_0^\infty t^{x-1}(1+t)^{y-x-1}\exp(-zt)dt$ 
is the confluent hypergeometric function of the second kind 
and $\Gamma(x) \equiv \int_0^\infty t^{t-1}\exp(-t)dt$ is the Gamma function.
The electron mean free path is given by $\ell_e = 1/(n_n\sigma_{en})$, 
where we will take the momentum transfer cross section $\sigma_{en}$ 
to be $\sigma_{en} = 10^{-15}~{\rm cm^2}$ by assuming that ${\rm H}_2$ 
dominates the gas \citep{FP62,YSH+08}.
Equation~\eqref{eq:fe} assumes that $\ell_e$ (or $\sigma_{en}$) is 
independent of $v_e$, which is a good assumption when the electron energy
is less than $10~{\rm eV}$ \citep[see, e.g.,  Figure~2 of][]{FP62}.  
In the limit of weak electric fields ($E \ll \Ecrit$), 
$f_{e0}$ reduces to the familiar Maxwell distribution 
\beq
f_{e0}^{\rm (M)} = \pfrac{m_e}{2\pi\kB T}^{3/2}\exp\left(-\frac{\eps_e}{\kB T}\right).
\label{eq:fe0_M}
\eeq
In the opposite limit ($E \gg \Ecrit$), 
$f_{e0}$ reduces to the Druyvesteyn distribution \citep{DP40}
\beq
f_{e0}^{\rm (D)} = \frac{1}{\pi \Gamma(\frac{3}{4})}\pfrac{3m_e^3}{4m_n(eE\ell_e)^2}^{3/4}
\exp\left(-\frac{3m_e \eps_e^2}{m_n(eE\ell_e)^2}\right).
\label{eq:fe0_D}
\eeq
In Figure~\ref{fig:DF}, we plot $f_{e0}$ for $E = 0$ and $100\Ecrit$ as a function of $\eps_e/\kB T$.

\begin{figure}[t]
\epsscale{1.1}
\plotone{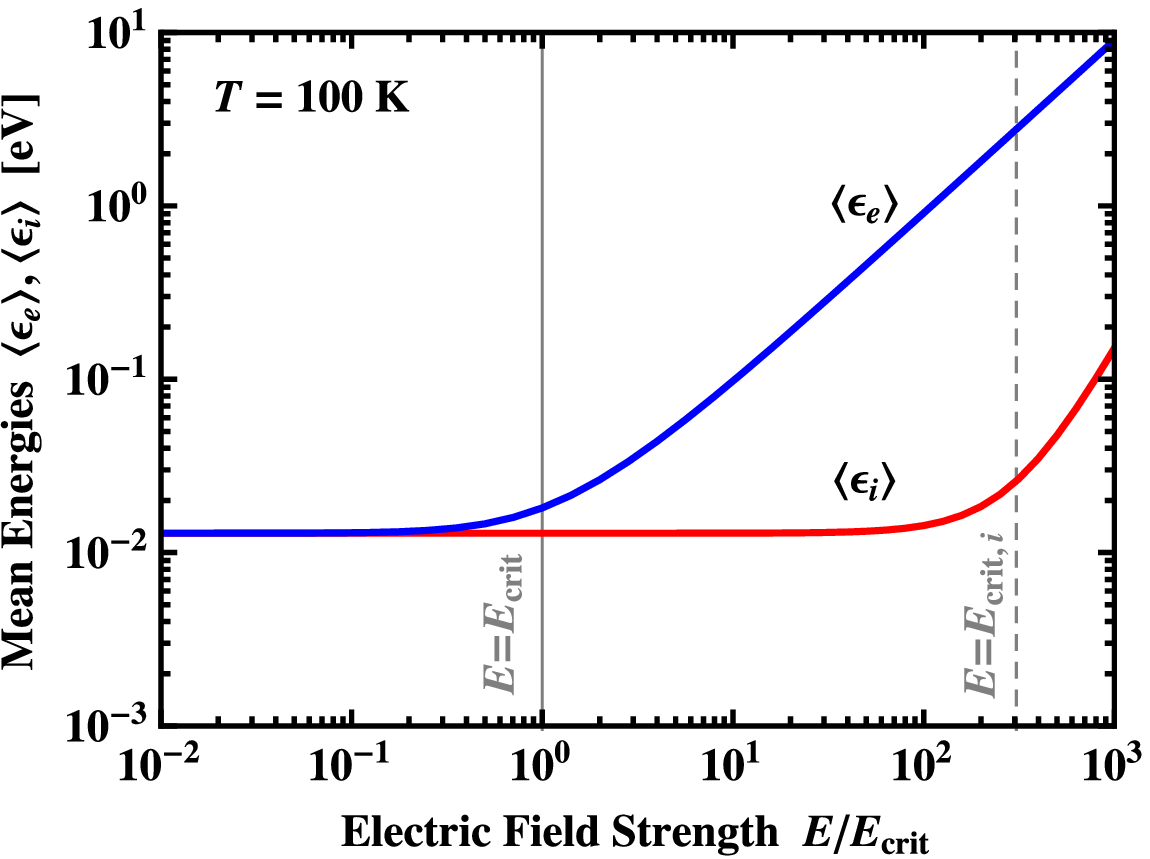}
\caption{Mean kinetic energies of electrons and ions, 
$\brac{\eps_e}$ (Equation~\eqref{eq:ee_exact}) and $\brac{\eps_i}$ (Equation~\eqref{eq:ei}), 
as a function of the normalized field strength $E/\Ecrit$.
The neutral gas temperature $T$ is assumed to be  $100~{\rm K}$. 
The solid and dashed vertical lines mark $E = \Ecrit$ and 
$E = \Ecriti$ (Equation~\eqref{eq:Ecriti}), respectively.
}
\label{fig:e}
\end{figure}

Substituting {Equations~\eqref{eq:fe}} and~\eqref{eq:fe0} 
into Equations~\eqref{eq:valpha||_def} and \eqref{eq:ealpha_def},
the mean velocity and energy of electrons are analytically obtained as
\beq
\brac{{\bm v}_{e||}} 
= -\frac{\Gamma(1+\chi,\chi)\exp\chi}{W(\chi)}
\frac{e{\bm E}\ell_e}{3\kB T} \sqrt{\dfrac{8 \kB T}{\pi m_e}},
\label{eq:ve||exact}
\eeq
{and}
\beq
\brac{\eps_e} 
= \frac{\chi U (\case{5}{2},\case{7}{2}+\chi,\chi )}
{U(\case{3}{2},\case{5}{2}+\chi,\chi )}
\frac{3\kB T}{2},
\label{eq:ee_exact}
\eeq
where $\Gamma(x,z) \equiv \int_z^\infty t^{x-1}\exp(-t)dt$ 
is the incomplete Gamma function. 
Figure~\ref{fig:e} plots Equation~\eqref{eq:ee_exact} as a function of $E/\Ecrit$ 
for $T = 100~{\rm K}$.

For later convenience, let us see how $\brac{{\bm v}_{e||}}$ and $\brac{\eps_e}$ behave 
in the limits of weak and strong electric fields.
Substituting Equations~\eqref{eq:fe0_M} and \eqref{eq:fe0_D}
 into Equations~\eqref{eq:valpha||_def} and \eqref{eq:ealpha_def}, we have
\beq
\brac{{\bm v}_{e||}}  \approx \left\{ \begin{array}{ll}
-\dfrac{e{\bm E}\ell_e}{3\kB T}\sqrt{\dfrac{8 \kB T}{\pi m_e}}, & E \ll \Ecrit, \\[4mm]
-\dfrac{\sqrt{2\pi}}{3^{3/4}\Gamma(\frac{3}{4})}
\dfrac{\sqrt{e E \ell_e}~}{(m_em_n)^{1/4}}\hat{{\bm E}}, & E \gg \Ecrit,
\end{array}\right.
\label{eq:ve||}
\eeq
\beq
\brac{\eps_e}  \approx \left\{ \begin{array}{ll}
\dfrac{3\kB T}{2}, & E \ll \Ecrit, \\[3mm]
\dfrac{\Gamma(\frac{5}{4})}{\Gamma(\frac{3}{4})}
\sqrt{\dfrac{m_n}{3m_e}} eE\ell_e, 
& E \gg \Ecrit.
\end{array}\right.
\label{eq:ee}
\eeq
We can see three important properties of the electron velocity distribution in the strong field limit. 
First, the drift speed $|\brac{{\bm v_{e||}}}|$ is proportional to $\sqrt{E}$, not to $E$.  
The reason is that the mean free time $\Delta t_e \sim \ell_e/\sqrt{\brac{v_e^2}}$
is inversely proportional to $\sqrt{E}$ in the strong field limit 
(this can be clearly seen by looking at the momentum conservation law 
of electrons; see Appendix~\ref{sec:kinetics}).
We will see in the following section that the nonlinearity of Ohm's law partly comes 
from the nonlinearity of $|\brac{{\bm v_{e||}}}|$.
Second, the mean electron energy is approximately given by $\brac{\eps_e} \approx 1.04(E/\Ecrit)\kB T
\approx (E/\Ecrit)\kB T$. 
Thus, if $T \sim 100~{\rm K}$, a field of  $E \approx 100\Ecrit$ gives a mean electron energy of
$\brac{\eps_e} \sim 1~{\rm eV}$ (see also Figure~\ref{fig:e}).
Third, the kinetic energy associated with the drift motion, 
 $m_e\brac{{\bm v_{e||}}}^2/2$, is smaller than the total kinetic energy 
$\brac{\eps_e}$ by the factor $(m_e/m_n)^{1/2} \sim 0.01$. 
This means that electrons's random motion dominates over 
systematic motion even in the strong field limit.
Thus, in weakly ionized plasmas, electric fields ``heat'' rather than ``accelerate'' electrons.

\subsubsection{Ions}\label{sec:ion}
Unlike for electrons, there is no closed expression for the 
velocity distribution function of ions at high electric fields. 
The difference arises from the fact that ion's momentum transfer cross section depends on
the ion--neutral collision velocity (instead, the mean collision time is approximately constant) 
owing to the polarization force between ions and neutrals \citep{W53}.
For this reason, we approximate $f_i$ by the offset Maxwell distribution \citep{H39}
\beq
f_i({\bm E},{\bm v}_i) = \pfrac{m_i}{2\pi \kB T_i}^{3/2}
\exp\biggl( -\frac{m_i({\bm v}_i-\brac{{\bm v}_{i||}})^2}{2\kB T_i}\biggr),
\label{eq:fi}
\eeq
where the ion drift velocity $\brac{{\bm v}_{i||}}$ and ion temperature $T_i$ are given by
\beq
\brac{{\bm v}_{i||}} = \frac{m_i+m_n}{m_im_n}e{\bm E}\Delta t_i,
\label{eq:vi||}
\eeq
\beq
\frac{3}{2}\kB T_i  
= \frac{3}{2}\kB T + \frac{1}{2}m_n\brac{{\bm v}_{i||}}^2,
\label{eq:Ti}
\eeq
respectively. 
The mean free time $\Delta t_i$ is the inverse of the frequency of collisions with neutrals,
and is given by $\Delta t_i = 1/K_{in}n_n$, where $K_{in}$ is the momentum transfer
rate coefficient for ion--neutral collisions (assumed to be a constant).  
We take $K_{in} = 1.6\times 10^{-9}~{\rm cm^3~s^{-1}}$ following \citet{NU86a}.
From Equation~\eqref{eq:ealpha_def}, the mean kinetic energy is 
\beqn
\brac{\eps_i} 
&=&  \frac{3}{2}\kB T_i + \frac{1}{2}m_i \brac{{\bm v}_{i||}}^2 
\nonumber \\
&=&
\frac{3}{2}\kB T + \frac{1}{2}(m_n+m_i)\brac{{\bm v_{i||}}}^2
\nonumber \\
&=&
\frac{3}{2}\kB T +  \frac{(m_n+m_i)^3(eE\Delta t_i)^2}{2(m_im_n)^2}.
\label{eq:ei}
\eeqn
Although the distribution function given by Equation~\eqref{eq:fi} is approximate, 
Equations~\eqref{eq:vi||} and \eqref{eq:ei} are the {\it exact} 
expressions for $\brac{{\bm v}_{i||}}$ and $\brac{\eps_i}$
of ions having a constant mean free time \citep[][see also Appendix~\ref{sec:kinetics}]{W53}. 

As we will show below, electric heating is much less efficient for ions than for electrons.
In the second (or third) line of Equation~\eqref{eq:ei}, 
the first and second terms account for heating by neutrals and electric fields, respectively. 
The second term dominates when $E > E_{{\rm crit},i}$, where
the threshold field strength $E_{{\rm crit},i}$ is given by
\beq
E_{{\rm crit},i} \equiv  \frac{m_i m_n \sqrt{3\kB T}}{(m_i+m_n)^{3/2}e\Delta t_i}. 
\label{eq:Ecriti}
\eeq
However, $\Ecriti$ is much larger than $\Ecrit$ because
\beqn
\frac{\Ecriti}{\Ecrit}&=& \sqrt{\frac{m_n}{2m_e\kB T}}
\frac{K_{in} m_i m_n}{\sigma_{en} (m_i+m_n)^{3/2}} \nonumber \\
&\sim& 300\pfrac{m_i}{30~{\rm amu}}^{-1/2}\pfrac{T}{100~{\rm K}}^{-1/2},
\label{eq:Ecritratio}
\eeqn
where we have assumed $m_i \gg m_n$, as is the case 
for dominant ions in protoplanetary disks like ${\rm HCO}^+$. 
Therefore, for $T \sim 10$--$1000~{\rm K}$,
ion heating {becomes} significant only at $E \ga 100$--$1000\Ecrit$.
As an example, in Figure~\ref{fig:e}, we compare $\brac{\eps_i}$ with $\brac{\eps_e}$ 
at $T = 100~{\rm K}$.
We see that ions start to be heated up only after electrons are heated to $\sim 1~{\rm eV}$.

\subsection{Charge Reactions}\label{sec:reaction}

We consider two ionizing mechanisms. 
One is the conventional ``external'' ionization by high-energy particles.
The sources may include galactic cosmic rays \citep{UN81}, 
stellar X-rays \citep{GNI97} and FUV \citep{PC11b}, 
and/or $\gamma$ rays from radionuclides \citep{UN09}. 
This process is characterized by a constant ionization rate $\zeta$ 
(the rate at which a single neutral gas particle is ionized).
The second mechanisms is impact ionization by electrically heated electrons. 
This is an ``internal'' ionization process in the sense 
that its rate is proportional to the electron number density $n_e$. 
Its rate also depends on the energy distribution of the electrons, 
and consequently on the strength $E$ of the applied electric field
 (see Section~\ref{sec:Kstar}). 
We neglect impact ionization by ions since electrons are always hotter than ions (see Section~\ref{sec:ion}).
We also neglect thermal ionization by assuming that the temperature $T$ of the neutral gas is much lower than $1000~{\rm K}$ \citep{U83}. 
Secondary electron emission from dust grains is also neglected 
since it becomes important only when the electron energy is above $100~{\rm eV}$ \citep{CMR93,WHR95}.
Photoelectric emission from grains can become important when 
strong UV irradiation is present \citep{S41,WD01b}, but we do not consider this in this study.

Ionized particles are removed from the gas through 
gas-phase recombination and sticking to dust grains.
By the latter process, dust grains on average obtain a negative charge  
because electrons have a higher random velocity than ions. 
In this study, we express the mean charge of the grains by $eZ$, where $Z <0$.
We will also express the mean charge in terms of the grain surface potential 
\beq
\phi = \frac{eZ}{a},
\label{eq:phi_def}
\eeq
where $a$ is the grain radius.
Because of Coulomb interaction, the plasma accretion rates of the grains depend not only on $E$
but also on $\phi$ (see Section~\ref{sec:Kdalpha}).   

The charge reactions mentioned above determine how $n_i$, $n_e$, and $Z$ evolve with time $t$. 
This is described by the rate equations
\beq
\frac{dn_i}{dt} = \zeta n_n - K_{di}(E,\phi)n_dn_i - K_{\rm rec}(E) n_i n_e + K_*(E)n_nn_e,
\label{eq:evol_ni}
\eeq
\beq
\frac{dn_e}{dt} = \zeta n_n - K_{de}(E,\phi)n_dn_e - K_{\rm rec}(E) n_i n_e + K_*(E)n_nn_e,
\label{eq:evol_ne}
\eeq
\beq
\frac{dZ}{dt} = K_{di}(E,\phi)n_i - K_{de}(E,\phi)n_e,
\label{eq:evol_Z}
\eeq
where 
$K_{d\alpha}~(\alpha=i,e)$, $K_{\rm rec}$, and $K_*$ are
the rate coefficients for plasma accretion by grains,
gas-phase recombination, and impact ionization by electrons, respectively. 
Plasma heating affects the solution of these equations through 
the $E$ dependences of the rate coefficients.

The set of Equations~\eqref{eq:evol_ni}--\eqref{eq:evol_Z} 
has a constant of integration, $\rho_c \equiv e(n_i - n_e + Zn_d)$, 
which is the net charge of the gas--dust mixture.
In this study, we assume $\rho_c = 0$ and obtain the charge neutrality condition
\beq
n_i - n_e + Zn_d = 0.
\label{eq:neut}
\eeq
It is important to note here that the plasma gas is generally nonneutral, 
i.e., $n_i \not= n_e$, because the grains contribute to the overall charge
neutrality of the gas--dust mixture. 
This is particularly true when the ionization rate is low and/or small dust grains are abundant.

\subsubsection{Gas-Phase Recombination}\label{sec:Krec}
Gas-phase recombination is dissociative for molecular ions like ${\rm HCO^+}$.
For ${\rm HCO^+}$, \citet{GBJD88} provide an empirical fit to 
the experimental data of the recombination rate coefficient
\beq
K_{\rm rec} = 2.4 \times 10^{-7} 
\pfrac{T_e}{300~{\rm K}}^{-0.69}~{\rm cm^3~s^{-1}},
\label{eq:Krec}
\eeq
where $T_e$ is the electron temperature. 
In this study, we use Equation~\eqref{eq:Krec} but we replace $T_e$ by $2\brac{\eps_e}/3\kB$.  
If metal ions like ${\rm Mg}^+$ are dominant, 
gas-phase recombination is radiative, and therefore $K_{\rm rec}$ becomes 
much lower than that given by Equation~\eqref{eq:Krec}. 
However, such a difference is unimportant 
when plasma accretion by dust grains dominates over gas-phase recombination.

\subsubsection{Plasma Sticking to Dust Grains}\label{sec:Kdalpha}
The rate coefficient for plasma accretion by grains 
is given by
\beq
K_{d\alpha}(E,\phi) = \int f_\alpha({\bm E}, {\bm v}_\alpha) 
\sigma_{d\alpha}(\eps_\alpha,\phi)v_\alpha d^3v_\alpha,
\label{eq:Kdalpha_def}
\eeq
where $\sigma_{d\alpha}$ is the effective collision cross section.
In this study, we adopt $\sigma_{d\alpha}$ of the form \citep{S41,SM02}
\beq
\sigma_{d\alpha}(\eps_\alpha,\phi) = \left\{ \begin{array}{ll} 
\pi a^2 \left(1-\dfrac{q_\alpha\phi}{\eps_\alpha}\right), 
& \eps_\alpha > q_\alpha \phi, \\[1mm]
0, & \eps_\alpha < q_\alpha \phi.
\end{array}\right.
\label{eq:sigma_alpha}
\eeq
It should be noted that the above expression assumes
that plasma particles perfectly stick to grains upon a collision.
Some ionization models in the astronomical literature \citep[e.g.,][]{U83,NNU91,IN06a,B11a} 
assume that the electron sticking probability rapidly decreases 
as the electron energy increases beyond $\sim 100~{\rm K} \sim 10^{-2}~{\rm eV}$.
However, such an assumption is inconsistent with the results of laboratory experiments. 
There is ample evidence that dust grains in plasmas are highly negatively charged even
 if the electron temperature is as high as  
  $0.1$--$10~{\rm eV}$ \citep[e.g.,][]{MTP94,BDM94,WHR95,RKZ+04},
which is well reproduced by models assuming perfect sticking \citep{KRZ+05}.
Therefore, perfect sticking is a more natural assumption 
as long as the secondary electron emission from dust grains is negligible (i.e., $\eps_e \la 100~{\rm eV}$).
Equation~\eqref{eq:sigma_alpha} also neglects 
the polarization force between grains and charged particles, 
which is valid for $aT \ga 10~\micron~{\rm K}$ \citep{DS87}.
 
\begin{figure}[t]
\epsscale{1.1}
\plotone{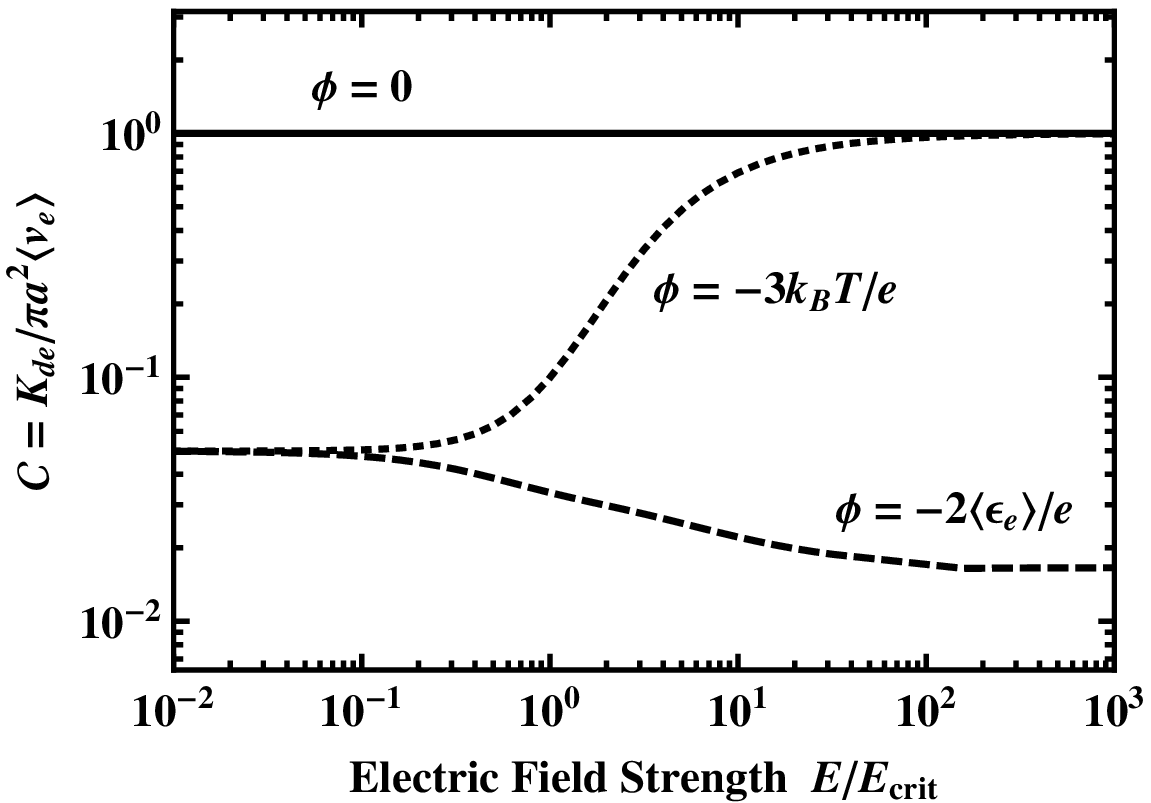}
\caption{Coulomb reduction factor ${\cal C} = K_{de}/\pi a^2 \brac{v_e}$ (Equation~\eqref{eq:C_def})
as a function of $E/\Ecrit$ for $\phi = 0$ (solid curve), 
$\phi = -3\kB T/e$ (dotted curve), and $\phi = -2\brac{\eps_e}/e$ (dashed curve).
}
\label{fig:Kde}
\end{figure}

For electrons, we use   
Equation~\eqref{eq:fe} and obtain
\beqn
K_{de} &=& 
\pi a^2 \sqrt{\dfrac{8 \kB T}{\pi m_e}} \frac{1}{W(\chi)}
\bigl[\left(\psi+\chi\right)^{1+\chi}\exp\left(-\psi \right) 
\nonumber \\
&& +\left(1-\psi\right)\Gamma\left(1+\chi,\psi+\chi\right)
\exp\chi \bigr],
\label{eq:Kde_exact}
\eeqn
where  $\psi \equiv  -e \phi/\kB T$ is the negative surface potential of the grains normalized by $\kB T$
(note that we assume $\phi < 0$ and hence $\psi>0$). 
In the absence of photoelectric and secondary electron emissions, 
dust grains tend to be negatively charged  because electrons move much faster than ions. 
For neutral grains ($\phi \to 0$), $K_{de}$ is simply given by 
the product of grain's geometric cross section and electron's mean speed,
\beq
K_{de}(E,\phi=0) = \pi a^2 \brac{v_e},
\eeq 
with
\beqn
\brac{v_e} &\equiv& \int |{\bm v}_e| f_{e0} d^3v_e  \nonumber \\
&=& \sqrt{\dfrac{8 \kB T}{\pi m_e}} \frac{1}{W(\chi)}
 \bigl[ \chi^{1+\chi} +\Gamma\left(1+\chi,\chi\right)
\exp\chi \bigr].
\label{eq:ve_exact}
\eeqn
Therefore, 
the dimensionless quantity
\beq
{\cal C}(E,\phi) \equiv \frac{K_{de}}{\pi a^2 \brac{v_e}}
\label{eq:C_def}
\eeq
measures how much the electron--grain collision rate is reduced by the Coulomb repulsion.
We will call ${\cal C}$ the Coulomb reduction factor fo electron--grain collisions.
By using $f_{e0}^{\rm(M)}$ and $f_{e0}^{\rm(D)}$ instead of Equations~\eqref{eq:fe},
one can obtain the asymptotic expressions of $\brac{v_e}$ and ${\cal C}$ in the weak and strong field limits,
\beq
\brac{v_e} 
\approx \left\{ \begin{array}{ll}
\sqrt{\dfrac{8 \kB T}{\pi m_e}}, & E \ll \Ecrit, \\[3mm]
\dfrac{\sqrt{2}}{3^{1/4}\Gamma(\frac{3}{4})}
\pfrac{m_n}{m_e}^{1/4} \sqrt{\dfrac{eE\ell_e}{m_e}}, ~ & E \gg \Ecrit,
\end{array}\right.
\label{eq:ve}
\eeq
and
\beq
{\cal C}  \approx \left\{ \begin{array}{ll}
\exp \left(\dfrac{-e|\phi|}{\kB T}\right),  & E \ll \Ecrit, \\[4mm]
\ \exp(-X^2) - \sqrt{\pi}X \, {\rm erfc}(X),   & E \gg \Ecrit,
\end{array}\right.
\label{eq:calC}
\eeq
where  ${\rm erfc}(x)$ is the complementary error function and
\beq
X \equiv \sqrt{\dfrac{3m_e}{m_n}}\dfrac{|\phi|}{E\ell_e}. 
\eeq
The form of ${\cal C}$ in the weak field limit is well known \citep{S41,SM02}.
{
Because $e|\phi|/\kB T \approx 1.5 e|\phi|/\brac{\eps_e}$ for $E \ll \Ecrit$ 
and $X \approx 0.74 e|\phi|/\brac{\eps_e}$ for $E \gg \Ecrit$,
Equation~\eqref{eq:calC} indicates that ${\cal C}$ is determined by the ratio 
between the electric and kinetic energies $e|\phi|/\brac{\eps_e}$. 
This can also be seen in Figure~\ref{fig:Kde}, 
where we plot the exact form of ${\cal C}$ 
as a function of $E/\Ecrit$ for three cases $e\phi = 0$, $-3\kB T$, and $-2\brac{\eps_e}$.
As we see, ${\cal C}$ is nearly constant for $\phi \propto \brac{\eps_e}$,
while ${\cal C}$ increases toward unity for constant $\phi$. 
}

For ions, we use Equation~\eqref{eq:fi} and obtain
\beqn
K_{di} &=& \pi a^2 \Biggl[ \sqrt{\dfrac{2 \kB T_i}{\pi m_i}}
\exp\left(-\frac{m_i\brac{{\bm v}_{i||}}^2}{2 \kB T_i} \right)  
\nonumber \\
&&+ |\brac{{\bm v}_{i||}}| 
\left( 1 + \frac{\kB T_i + 2e|\phi| }{m_i \brac{{\bm v}_{i||}}^2} \right) {\rm erf}
\left( \sqrt{\frac{m_i}{2\kB T_i}}|\brac{{\bm v}_{i||}}| \right) \Biggr],
\label{eq:Kdi_exact}
\eeqn
where
${\rm erf}(x)$ is the error function.
In the limits of $E\ll\Ecriti$ and $E\gg\Ecriti$, Equation~\eqref{eq:Kdi_exact} reduces to
\beq
K_{di}  \approx \left\{ \begin{array}{ll}
\pi a^2 \sqrt{\dfrac{8 \kB T}{\pi m_i}}\left(1 + \dfrac{e|\phi|}{\kB T} \right),  & E \ll \Ecriti, \\[3mm]
\pi a^2 |\brac{{\bm v}_{i||}}| \left(1 + \dfrac{2e|\phi|}{m_i\brac{{\bm v}_{i||}}^2} \right), 
& E \gg \Ecriti,
\end{array}\right.
\label{eq:Kdi}
\eeq
where we have used $m_i \gg m_n$ for the high-field expression.

\subsubsection{Impact Ionization}\label{sec:Kstar}
The impact ionization rate coefficient $K_*$ is given by
\beq
K_{*}(E) = \int  f_e({\bm E},{\bm v}_e) \sigma_*(\eps_e) v_e d^3 v_e,
\label{eq:Kstar_def}
\eeq
where $\sigma_*$ is the impact ionization cross section of neutrals.
For $\sigma_*$, we adopt Thomson's expression \citep{T12}
\beq
\sigma_*(\eps_e) =  \left\{ \begin{array}{ll}
\dfrac{\pi N_* e^4}{\eps_e^2} \left(\dfrac{\eps_e}{\IP}-1\right), & \eps_e > \IP, \\
0, & \eps_e < \IP,
\end{array}\right.
\label{eq:sigma_star}
\eeq
where $N_*$ is the number of bound electrons in the outermost shell of the neutrals
and $\IP$ is the ionization potential of the outermost bound electrons.
Equation~\eqref{eq:sigma_star} well approximates 
experimentally obtained ionization cross sections
unless $\eps_e$ is much larger than $\IP$ \citep{L67}.
We only consider the impact ionization of ${\rm H_2}$ molecules 
(${\rm IP} = 15.4~{\rm eV}$) because they dominate the gas of 
protoplanetary disks.
 However, if there are a considerable number of metal atoms having a low ionization energy (e.g., K and Ca) 
 in the gas phase,  they would effectively lower the value of IP in Equation~\eqref{eq:sigma_star}. 

Since impact ionization is only important for  $E \gg \Ecrit$, 
it is sufficient to evaluate $K_*$ using the Druyvesteyn distribution $f_{e0}^{\rm(D)}$ (Equation~\eqref{eq:fe0_D}).
This allows us to analytically perform the integration in Equation~\eqref{eq:Kstar_def}, yielding 
\beq
K_*(E) = \frac{\sqrt{2}\pi^{3/2} N_*e^4}{\Gamma(\frac{3}{4})\sqrt{m_e}\,\IP^{3/2}}
 \sqrt{Y} \left( 
{\rm erfc}(Y) + \frac{Y}{\sqrt{\pi}}{\rm Ei}(-Y^2) \right)
 \nonumber \\ 
\label{eq:Kstar}
\eeq
with
\beq
Y \equiv \sqrt{\frac{3m_e}{m_n}}\frac{\IP}{eE\ell_e},
\eeq
where 
${\rm Ei}(x) \equiv -\int_{-x}^\infty t^{-1}\exp(-t) dt$ is the exponential integral.
Note that $K_*$ is determined by the ratio $\brac{\eps_e}/{\rm IP}$
because  $Y \approx 0.74 \IP/ \brac{\eps_e}$.
Since the Druyvesteyn distribution neglects inelastic losses, 
Equation~\eqref{eq:Kstar} must be taken as a very crude estimate for $K_*$.

\begin{figure}[t]
\epsscale{1.1}
\plotone{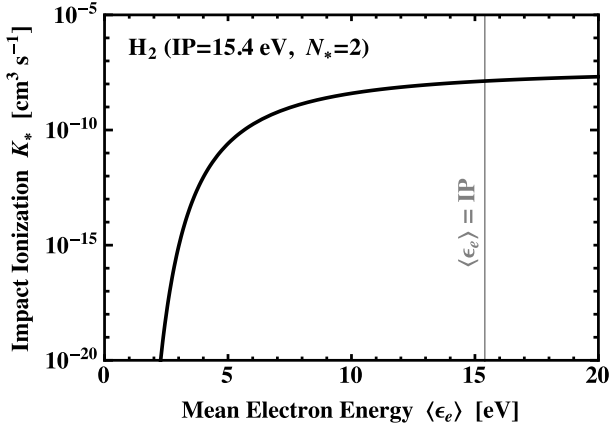}
\caption{Impact ionization rate coefficient $K_*$ (Equation~\eqref{eq:Kstar}) for ${\rm H_2}$ 
as a function of the mean electron energy $\brac{\eps_e}$.
The vertical line marks $\brac{\eps_e} = {\rm IP}$.
}
\label{fig:Kstar}
\end{figure}
Figure~\ref{fig:Kstar} shows $K_*$ for hydrogen molecules as a function of  $\brac{\eps_e}$.
As we can see, $K_*$ abruptly increases 
before $\brac{\eps_e}$ reaches the ionization threshold ${\rm IP} = 15.4~{\rm eV}$.
This means that electrons at the high-energy tail of the energy distribution significantly 
contribute to the impact ionization.
This would remain true, at least qualitatively, even if inelastic ionization losses are included.

\subsubsection{Charge Equilibrium Solution}\label{sec:steady}
In this study, we follow \citet{O09} and calculate the equilibrium solutions 
to the rate equations in an analytic way. 
First we solve Equations~\eqref{eq:evol_ni} and \eqref{eq:evol_ne} 
with respect to $n_i$ and $n_e$ under the equilibrium condition $dn_i/dt = dn_e/dt = 0$. 
The solution $n_\alpha \equiv n_\alpha^{\rm(eq)}$ $(\alpha=i,e)$ is then a function of $E$ and $\phi$.
It is easy to show that the solution is given by
\beq
n_\alpha^{(\rm eq)}(E, \phi) = \frac{\zeta n_n}{K_{d\alpha}(E,\phi)n_d}
\left( 
\sqrt{\frac{1}{\cal S} + \left(\frac{1-{\cal I}}{2}\right)^2} + \frac{1-{\cal I}}{2} 
\right)^{-1},
\label{eq:nalpha_eq}
\eeq
where the dimensionless quantities ${\cal S}$ and ${\cal I}$ are defined by
\beq
{\cal S}(E,\phi) = \frac{K_{di}(E,\phi)K_{de}(E,\phi)n_d^2}{K_{\rm rec}(E) \zeta n_n},
\label{eq:calS}
\eeq
\beq
{\cal I}(E,\phi) = \frac{K_*(E)n_n}{K_{de}(E,\phi)n_d}.
\label{eq:calI}
\eeq
The parameter ${\cal S}$ indicates which of gas-phase recombination and 
plasma sticking onto grains dominates (the latter dominates if ${\cal S} > 1$), 
while ${\cal I}$ indicates which of electron sticking onto grains and impact ionization dominates
(the latter dominates if ${\cal I} > 1$).
{In this paper, we will call ${\cal S}$ the {grain} recombination parameter.}

In order to determine $\phi$ as a function of $E$, we substitute Equation~\eqref{eq:nalpha_eq}  
into the charge neutrality condition (Equation~\eqref{eq:neut}) to obtain
\beq
n_i^{(\rm eq)}(E, \phi) - n_e^{(\rm eq)}(E, \phi) + \frac{a n_d}{e}\phi = 0,
\label{eq:master}
\eeq
where we have used Equation~\eqref{eq:phi_def} to rewrite $Z$ as $a\phi/e$.
Equations~\eqref{eq:nalpha_eq}--\eqref{eq:master} 
reduce to Equations~(27)--(30) of \citet{O09} in the limit $E \to 0$.
{As shown by \citet{O09}, one can extend our Equations~\eqref{eq:nalpha_eq}--\eqref{eq:master} for arbitrary grain size distribution $dn_d/da$ (number of grains per unit grain radius) 
if one replaces $an_d$ in Equation~\eqref{eq:master} with $\int a (dn_d/da) da$, 
and $\pi a^2 n_d$ in $K_{d\alpha}n_d$ with $\int \pi a^2 (dn_d/da) da$.}

{We solve Equation~\eqref{eq:master} with respect to $\phi$ using the Newton--Raphson method.
If impact ionization is neglected (${\cal I}=0$), the left-hand side of Equation~\eqref{eq:master} monotonically increases with $\phi$, 
so Equation~\eqref{eq:master} has only one root for each value of $E$. 
In this case, the Newton--Raphson procedure converges to the single root with an arbitrary initial guess. 
With impact ionization,  Equation~\eqref{eq:master} can possess three roots for a certain range of $E$
(see Section~\ref{sec:discharge}).
In this case, we search for all the roots by varying the initial guess gradually from $\phi = -0.1~{\rm V}$ to $-10~{\rm V}$.
}

\subsection{Model Parameters}
\begin{deluxetable}{lccc}
\tablecaption{Model Parameters}
\tablecolumns{4}
\tablewidth{0pt}
\tablehead{
\colhead{Model} &  \colhead{$\zeta~({\rm s^{-1}})$} & \colhead{$f_{dg}$} 
&  \colhead{Impact ionization?} 
}
\startdata
A & $10^{-17}$ &  $10^{-6}$ & No \\ 
B, \Bstar &  $10^{-17}$ &  $10^{-4}$ & No (B), Yes (\Bstar) \\ 
C, \Cstar &  $10^{-17}$ & $10^{-2}$ & No (C), Yes (\Cstar) \\
D & $10^{-19}$ &  $10^{-2}$ & No 
\enddata
\tablecomments{The other parameters are fixed to 
$m_n = 2.3~{\rm amu}$, $m_i = 29~{\rm amu}$,  
$T = 100~{\rm K}$, $n_n = 10^{12}~{\rm cm^{-3}}$, $\IP = 15.4~{\rm eV}$, 
$a = 1~\micron$, and $\rho_\bullet = 2~{\rm g~cm^{-3}}$.
}
\label{tab:param}
\end{deluxetable}
We consider four models (A, B, C, and D) without impact ionization 
and two models (\Bstar and \Cstar) with impact ionization.
The external ionization rate $\zeta$ and dust-to-gas mass ratio $f_{dg}$ 
for these models are listed in Table~\ref{tab:param}.
We fix the neutral mass $m_n = 2.3~{\rm amu}$, ion mass $m_i = 29~{\rm amu}$ (which is the mass of ${\rm HCO}^+$), neutral temperature $T = 100~{\rm K}$, neutral gas density $n_n = 10^{12}~{\rm cm^{-3}}$,  
ionization potential $\IP = 15.4~{\rm eV}$, grain size $a = 1~\micron$, and grain internal density $\rho_\bullet = 2~{\rm g~cm^{-3}}$.
The threshold field strengths for electron and ion heating are 
$\Ecrit = 1.1\times 10^{-9}~{\rm esu~cm^{-2}}$ and 
$\Ecriti = 3.3\times 10^{-7}~{\rm esu~cm^{-2}}$
{(in SI units, $\Ecrit = 3.3\times 10^{-5}~{\rm V~m^{-1}}$ 
and $\Ecrit = 1.0\times 10^{-2}~{\rm V~m^{-1}}$)}, respectively.
If we compare our choices of $T$ and $n_n$ with the minimum-mass solar nebula model \citep{H81},
we find that our models correspond to the disk midplane at 10 AU from the central star.

\section{Nonlinear Ohm's Laws without Impact Ionization}\label{sec:nodischarge}

Impact ionization is important  
only when the electric field strength $E$ is so high that the mean electron energy exceeds a few eV. 
However, heating of electrons still occurs  at lower $E$ 
and affects the rates of gas-phase recombination and electron sticking to dust grains.  
To isolate the role of plasma heating at relatively low $E$, 
we here ignore impact ionization (${\cal I} = 0$; models A, B, C, D) and focus on
how plasma heating changes the ionization state at lower $E$.  
The effect of impact ionization will be studied in Section~\ref{sec:discharge}.

\subsection{Ionization State}\label{sec:ionization}

\begin{figure*}
\epsscale{1.0}
\plotone{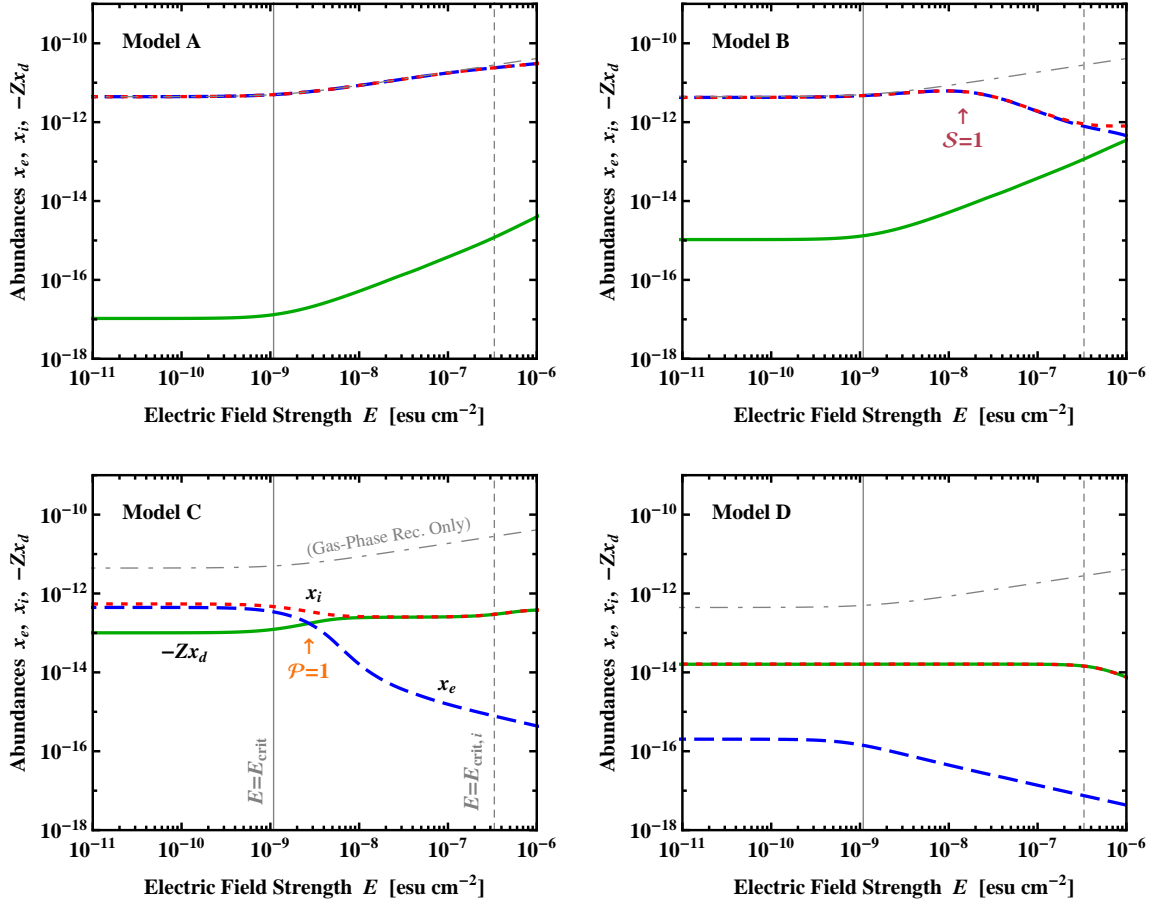}
\caption{Ion abundance $x_i$ (dotted curve), 
electron abundance $x_e$ (dashed curve), 
and grain charge abundance $-Zx_d$ (solid curve)
 as a function of the electric field strength $E$ for models A, B, C, and D.  
The dot-dashed curve shows $x_e$ and $x_i$
in the grain-free limit (${\cal S} \ll 1$; Equation~\eqref{eq:ne_S<1}).
The solid and dashed vertical
 lines mark $E = \Ecrit$ and $E = \Ecriti$, respectively. 
}
\label{fig:x}
\end{figure*}
\begin{figure}
\epsscale{1.1}
\plotone{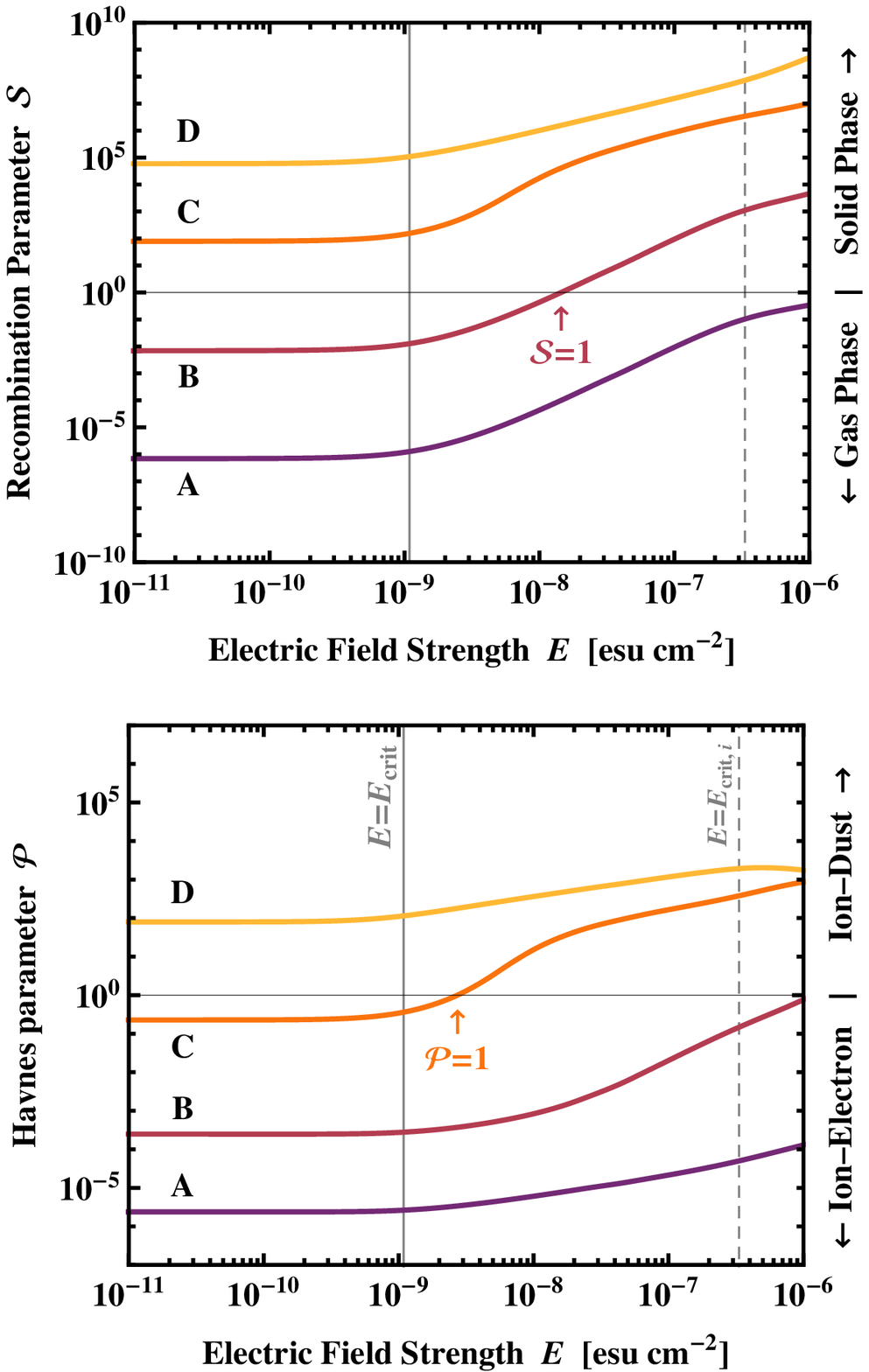}
\caption{
Lower panel: {grain} recombination parameter  ${\cal S}$ (Equation~\eqref{eq:calS}) 
as a function of the electric field strength $E$ for models A, B, C, and D. 
The solid and dashed vertical lines mark $E = \Ecrit$ and $E = \Ecriti$, respectively.
Gas-phase recombination dominates for ${\cal S} < 1$, 
while plasma sticking dominates for ${\cal S} > 1$.
Upper panel: Havnes parameter ${\cal P}$ (Equation~\eqref{eq:calP}) 
versus $E$ for the four models.
${\cal P} < 1$ corresponds to the ion--electron plasma state ($x_i \approx x_e$)
while ${\cal P} > 1$ to the ion--dust plasma state ($x_i \approx |Z|x_d$).
}
\label{fig:SP}
\end{figure}

Figure~\ref{fig:x} shows the ionization state of the four models as a function of $E$.
Here we plot the abundances of plasma particles, $x_e (= n_e/n_n)$ and $x_i (= n_i/n_n)$, 
and the negative grain charge $-Z$ in the abundance form $-Zx_d$ $(= -Zn_d/n_n)$.
From the charge neutrality, $x_i$ is equal to the sum of $x_e$ and $-Z x_d$. 
{Since impact ionization is not treated here, the equilibrium ionization state} 
is determined by the balance among external ionization, 
gas-phase recombination, and charging of dust grains.
This balance can be characterized by two dimensionless quantities.
The first one is ${\cal S}$ already introduced in Section~\ref{sec:steady} (Equation~\eqref{eq:calS}). 
This quantity is a diagnostic of the dominant recombination process: recombination mainly 
takes place in the gas phase for ${\cal S} <1$, and in the ``solid phase'' (i.e., on dust grains) 
for ${\cal S} >1$. 
The second one is given by
\beq
{\cal P} \equiv \frac{|Z|n_d}{n_e},
\label{eq:calP}
\eeq
which is known as the Havnes parameter \citep{H84} in the field of dusty plasma physics. 
This is a diagnostic of the charge neutrality in the gas--dust mixture. 
If ${\cal P} \ll 1$, gas-phase free electrons dominate over negatively charged grains, and 
the charge neutrality is approximately established within the gas phase (i.e., $n_i \approx n_e$).
If ${\cal P} >1$, negatively charged grains dominate, and the number of positive ions 
in the gas approximately balances with the number of free {electrons} on the grains $(n_i \approx -Z n_d)$.
The two dimensionless quantities measure how strongly dust grains affect the ionization state.

Figure \ref{fig:SP} plots these two diagnostics for the four models as a function of $E$. 
By definition, ${\cal S}$ and ${\cal P}$ increase 
as the amount of dust $f_{dg}$ is increased (see models A, B, and C).
They also increase as the ionization rate $\zeta$ is {\it decreased} (see models C and D)
because the presence of grains becomes more and more important as the ionized degree decreases. 
For fixed $f_{dg}$ and $\zeta$, ${\cal S}$ and ${\cal P}$ increase with $E$,
because more and more electrons are transferred from the gas to grains 
as the random velocity of electrons ($\approx$ collision velocity between electrons and grains) is increased.

\begin{figure}[t]
\epsscale{1.1}
\plotone{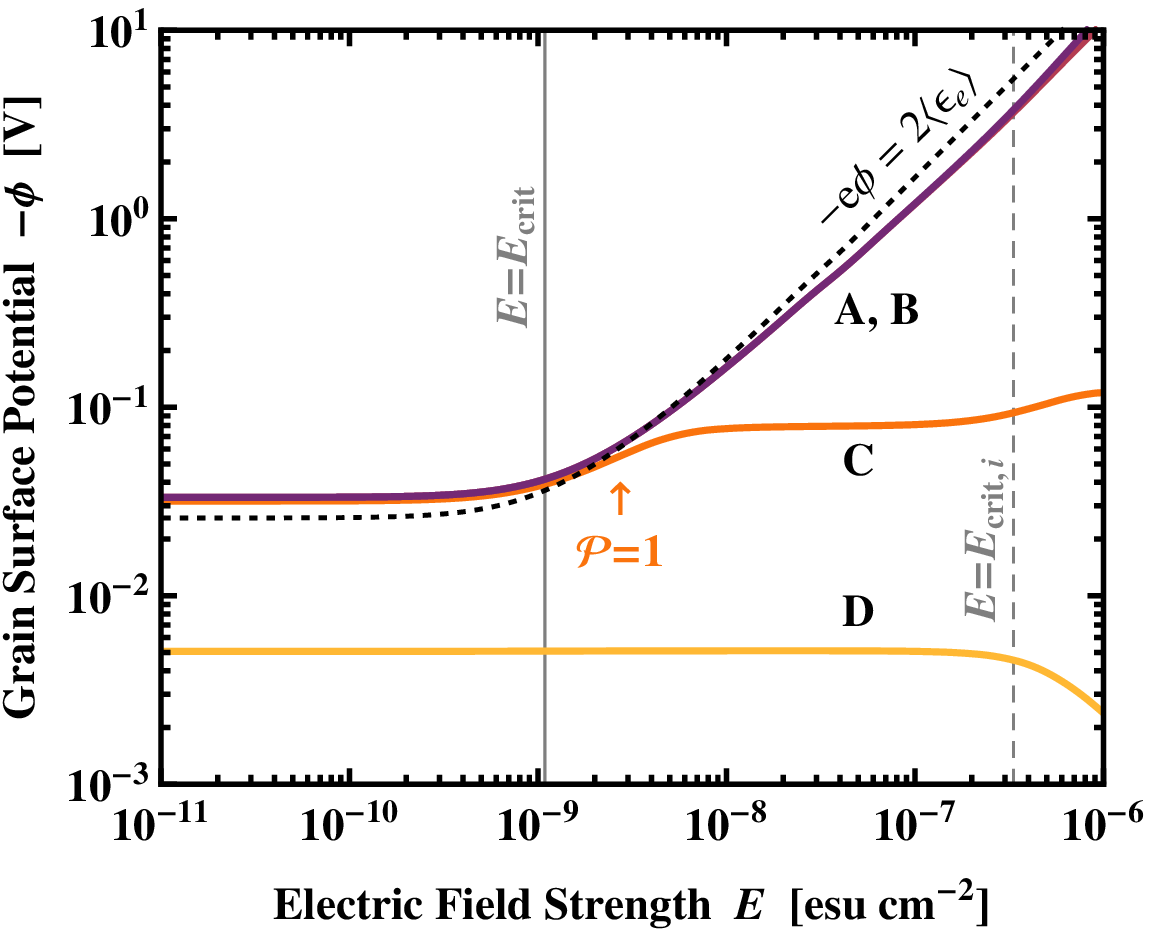}
\caption{Grain surface potential $-\phi = -eZ/a$ 
as a function of the electric field strength $E$ for models A, B, C, and D. 
The solid and dashed vertical lines mark $E = \Ecrit$ and $E = \Ecriti$, respectively. 
The dotted curve shows the relation $-e\phi = 2\brac{\eps_e}$.
}
\label{fig:phi}
\end{figure}

\begin{figure}[t]
\epsscale{1.1}
\plotone{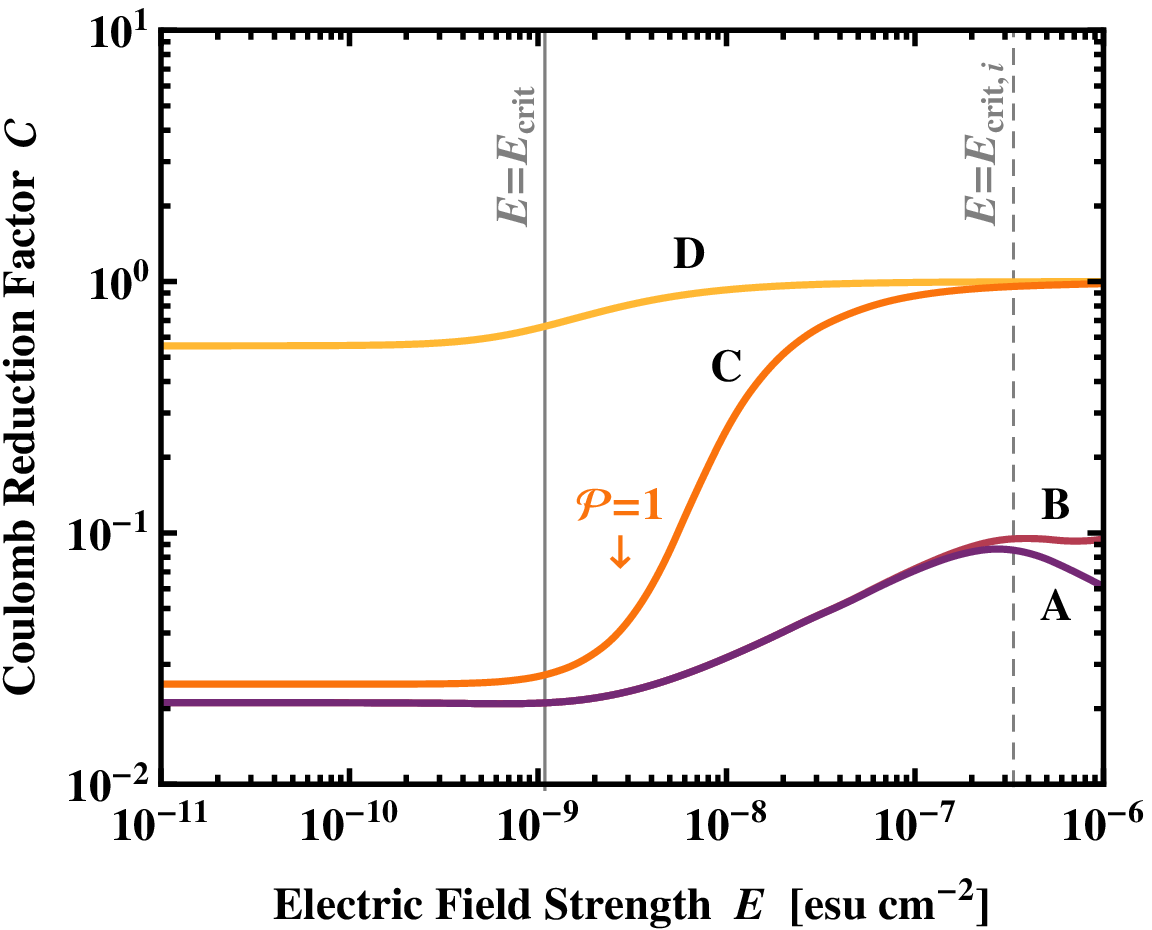}
\caption{Coulomb reduction factor ${\cal C}$ for grain--electron 
collisions (Equation~\eqref{eq:calC}) for models A, B, C, and D as a function of $E$. 
The solid and dashed vertical lines mark $E = \Ecrit$ and $E = \Ecriti$, respectively. 
}
\label{fig:C}
\end{figure}

Model A is characterized by the low dust-to-gas ratio $f_{dg} = 10^{-6}$ 
and is an example where ${\cal S} < 1$ and ${\cal P} < 1$ 
over the entire range of $E$ under consideration. 
In this model, the gas phase dominates both recombination and charge neutrality, 
and dust grains have essentially no effect on the ionization state of the gas.
The balance between external ionization rate $\zeta n_n$ and gas-phase recombination 
rate $K_{\rm rec}n_in_e$ gives an approximate expression for the plasma density $n_e$ $(\approx n_i)$
in the case of ${\cal S} \ll 1$, 
\beq
n_e \approx \sqrt{\frac{\zeta n_n}{K_{\rm rec}(E)}}.
\label{eq:ne_S<1}
\eeq
Note that $n_e$ increases with $E$ because the gas-phase recombination rate coefficient 
$K_{\rm rec}$ is a decreasing function of $\brac{\eps_e}$.
Since $n_e \approx n_i$, the electron flux $n_e\brac{v_e}$ is much higher than 
the ion flux $n_i \brac{v_i}$ (note that  in general $\brac{v_e} \gg \brac{v_i}$), 
so individual dust grains tend to be charged up 
so that Coulomb repulsion between the grains and electrons becomes effective. 
This can be seen in Figure~\ref{fig:phi}, where we plot $|\phi|$ as a function of $E$. 
In model A, $|\phi|$ increases with $E$ in the way that 
the relation $e|\phi| \sim 2 \brac{\eps_e}$ is approximately satisfied, i.e., 
in the way that the Coulomb repulsion energy between 
a grain and an electron upon collision is comparable to their collision energy. 
As a result, the Coulomb reduction factor ${\cal C}$ 
is much less than unity (${\cal C} <0.1$) for all $E$ as shown in Figure~\ref{fig:C}.

Model B is a more dusty case where $f_{dg}$ is 100 times larger than in model A. 
As a result, ${\cal S}$ now exceeds unity (i.e., 
solid-phase recombination becomes the dominant recombination process)
at $E \ga 10\Ecrit \approx 10^{-8}~{\rm esu~cm^{-2}}$.
In the case of ${\cal S} \gg 1$, $n_e$ is determined by the balance 
between external ionization rate $\zeta n_n$ and electron capture rate $K_{\rm rec}n_dn_e$, i.e., 
\beq
n_e \approx \frac{\zeta n_n}{K_{de}n_d} = \frac{\zeta n_n}{\pi a^2  n_d \brac{v_e}{\cal C}(E,\phi)}.
\label{eq:ne_S>1}
\eeq
Note that $n_e$ {\it decreases} with increasing $E$ 
because both $\brac{v_e}$ and ${\cal C}$ increase with $E$ (for ${\cal C}$, see Figure~\ref{fig:C}).
If ${\cal P} < 1$, as is the case in model B, $n_i$ is also given by Equation~\eqref{eq:ne_S>1}.

Model C is an even more dusty case where the dust-to-gas ratio is interstellar. 
In this model, ${\cal S} > 1$ over the entire range of $E$. 
In addition, at high $E$, ${\cal P}$ exceeds unity, 
i.e., dust grains become the dominant negative charge carriers.
We can see that $n_e$ rapidly decreases when ${\cal P}$ crosses unity. 
This is a positive feedback effect of grain's negative charging on electron depletion. 
As ${\cal P}$ exceeds $1$, 
$n_e$ becomes smaller than $n_i$, and 
the electron-to-ion flux ratio $n_e\brac{v_e}/n_i\brac{v_i}$ becomes closer to unity.
For this reason, individual dust grains tend to be less negatively charged than in the case of ${\cal P} \ll 1$. 
This can be seen in Figure~\ref{fig:phi}, where we see that 
$e |\phi|$ falls below $\brac{\eps_e}$ after ${\cal P}$ exceeds unity. 
As a result, the Coulomb repulsion between the grains and electrons become ineffective (${\cal C} \approx 1$),
leading to a further decrease in the electron number density according to Equation~\eqref{eq:ne_S>1}.  
Note that $n_i$ is constant at ${\cal P} > 1$ as long as ion heating is insignificant (i.e., $E \ll \Ecriti$).
This constant value is given by $n_i \approx \zeta n_n / \pi a^2 v_{i,T}$,
where $v_{i,T} = \sqrt{8 \kB T/\pi m_i}$ is the mean thermal speed of ions at $T_i = T$.

In model D, $\zeta$ is decreased by a factor of 100 from model C, 
and we see that ${\cal S} > 1$ and ${\cal P} > 1$ over the entire range of $E$.
The electron abundance decreases at $E > \Ecrit$, but more slowly than in model C 
since the Coulomb repulsion factor ${\cal C}$ 
is already close to unity from the beginning (see Figure~\ref{fig:C}).

\subsection{Current Density}\label{sec:current}
\begin{figure*}
\epsscale{1.0}
\plotone{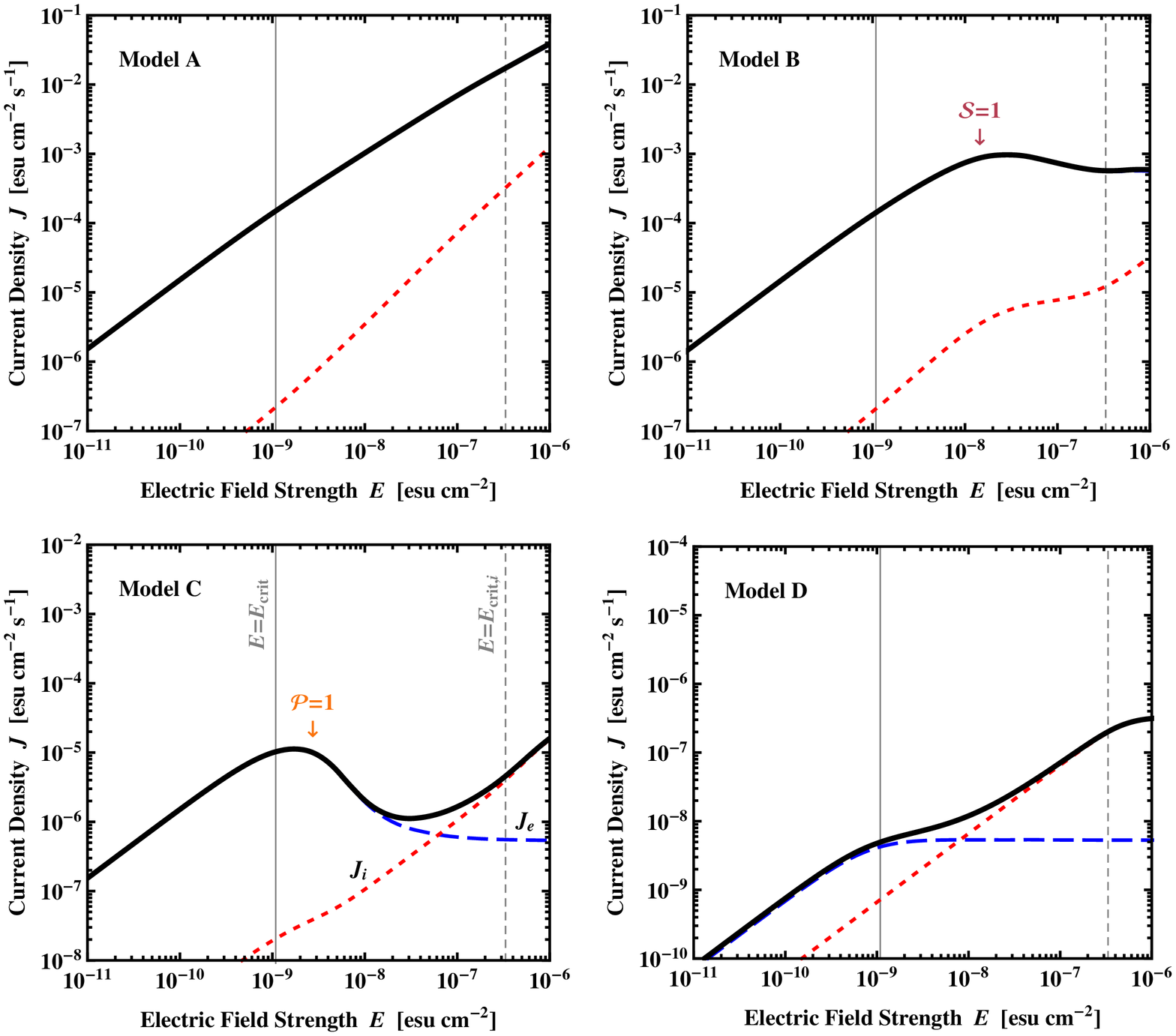}
\caption{Current density $J$ (solid curve) 
as a function of the electric field strength $E$ for models A, B, C, and D.
The dashed and dotted curves show the contributions 
from electrons and ions, $J_e$ and $J_i$, respectively.
For models A and B, the curves for $J$ and $J_e$ are indistinguishable.
The vertical lines mark $E = \Ecrit$.
}
\label{fig:J}
\end{figure*}
Figure~\ref{fig:J} show the magnitude of the current density 
$J = |{\bm J}|$ as a function of $E$ for models A, B, C, and D.
This is the sum of the ion and electron currents, 
$J_i = |{\bm J}_i| = e n_i |\brac{\bm v_{i||}}|$ and $J_e = |{\bm J}_e| = e n_e |\brac{\bm v_{e ||}}|$, 
which are also plotted in Figure~\ref{fig:J}.

As mentioned earlier, the $J$--$E$ relations are nonlinear in $E$ 
because the plasma heating changes $n_\alpha$ and $|\brac{\bm v_{\alpha||}}|$ ($\alpha = i, e$) 
at high $E$. 
The nonlinearity is, however, weak in model A. 
In this model, the electric current is dominated by $J_e$, 
which is proportional to the product of $n_e$ and $|\brac{{\bm v}_{e||}}|$. 
At $E \gg \Ecrit$, $|\brac{{\bm v}_{e||}}|$ increases more slowly than at $E \ll \Ecrit$
due to the enhanced frequency of electron--neutral collisions (see Equation~\eqref{eq:ve||}).
Meanwhile, $n_e$ increases with $E$ because the recombination rate coefficient $K_{\rm rec}$
 decreases with ${\cal \eps}_e$ (see Equations~\eqref{eq:Krec} and \eqref{eq:ne_S<1}).
These two opposing effects partially cancel out in the product $n_e|\brac{{\bm v}_{e||}}|$. 

In model B, $J$ behaves in the same way as in model A as long as ${\cal S}< 1$.   
The behavior changes when ${\cal S}$ crosses unity 
because $n_e$ becomes a decreasing function of $E$ (see Section~\ref{sec:ionization}). 
We see that the current  is approximately constant (precisely speaking, decreases very slowly with $E$)
at ${\cal S} > 1$.
This trend can be explained as follows.    
When ${\cal S} \gg 1$, $n_e$ is inversely proportional to the mean electron speed 
$\brac{v_e}$ (see Equation~\eqref{eq:ne_S>1}), which is, at $E \gg \Ecrit$, 
proportional to the electron drift speed $\brac{{\bm v}_{e||}}$.
Equations~\eqref{eq:ve||} and \eqref{eq:ve} imply that the ratio of the two velocities is 
\beq
|\brac{{\bm v}_{e||}}| \approx \sqrt{\frac{\pi m_e}{3m_n}}  \brac{v_e}. 
\label{eq:ve||ve}
\eeq
Therefore, in $J_e \propto n_e |\brac{{\bm v}_{e||}}|$, 
the dependence on $\brac{v_e}$ is canceled out, resulting in \beq
J_e \approx 
\frac{J_{e,\infty}}{{\cal C}},
\label{eq:Je_approx}
\eeq
where $J_{e,\infty}$ is a constant defined by
\beqn
J_{e,\infty} &\equiv& \sqrt{\frac{\pi m_e}{3m_n}} \frac{\zeta e n_n}{\pi a^2 n_d}
\nonumber \\
&\approx& 5 \times 10^{-5}\pfrac{10^{-4}}{f_{dg}}\pfrac{a}{1~\micron}
\pfrac{\zeta}{10^{-17}~{\rm s^{-1}}}~{\rm esu~cm^{-2}~s^{-1}}.~~
\eeqn
Hence, if ${\cal C}$ is independent of $E$, so is $J_e$.
As we have already seen in Section~\ref{sec:ionization},
${\cal C}$ varies only slowly with $E$ unless ${\cal P}$ crosses unity. 
This explains why  in model B the current is approximately constant at ${\cal S}>1$.

The result for model C is more complex, but can also be explained in a similar way.
We see that $J_e$ starts to decrease with $E$ at the point where ${\cal P}$ crosses unity. 
This is because the Coulomb reduction factor ${\cal C}$ increases from 0.025 to unity 
as ${\cal P}$ goes from $\ll 1$ to $\gg 1$.
This leads to a $40$-fold decrease in $J_e$ across ${\cal P} = 1$
as predicted by Equation~\eqref{eq:Je_approx}.
By contrast, the ion current $J_i$ 
continues increasing with $E$ 
because $n_i$ is approximately constant for ${\cal S}>1$. 
As a consequence, the net current $J = J_e + J_i$ forms an N-shaped curve in the $J$--$E$ diagram.  

In model D, $J_e$ immediately approaches $J_{e,\infty}$ at $E > \Ecrit$ 
since ${\cal C} \approx 1$ from the beginning.  
The net current monotonically increases with $E$ because of the presence of $J_i$.
At $E \ga \Ecriti$, $J_i$ also gets saturated at a constant value. 

\section{Nonlinear Ohm's Laws with Impact Ionization}~\label{sec:discharge}
We now include impact ionization (models \Bstar~and \Cstar) 
and see how it changes the ionization balance at very high electric field strengths.

\begin{figure*}
\epsscale{1.0}
\plotone{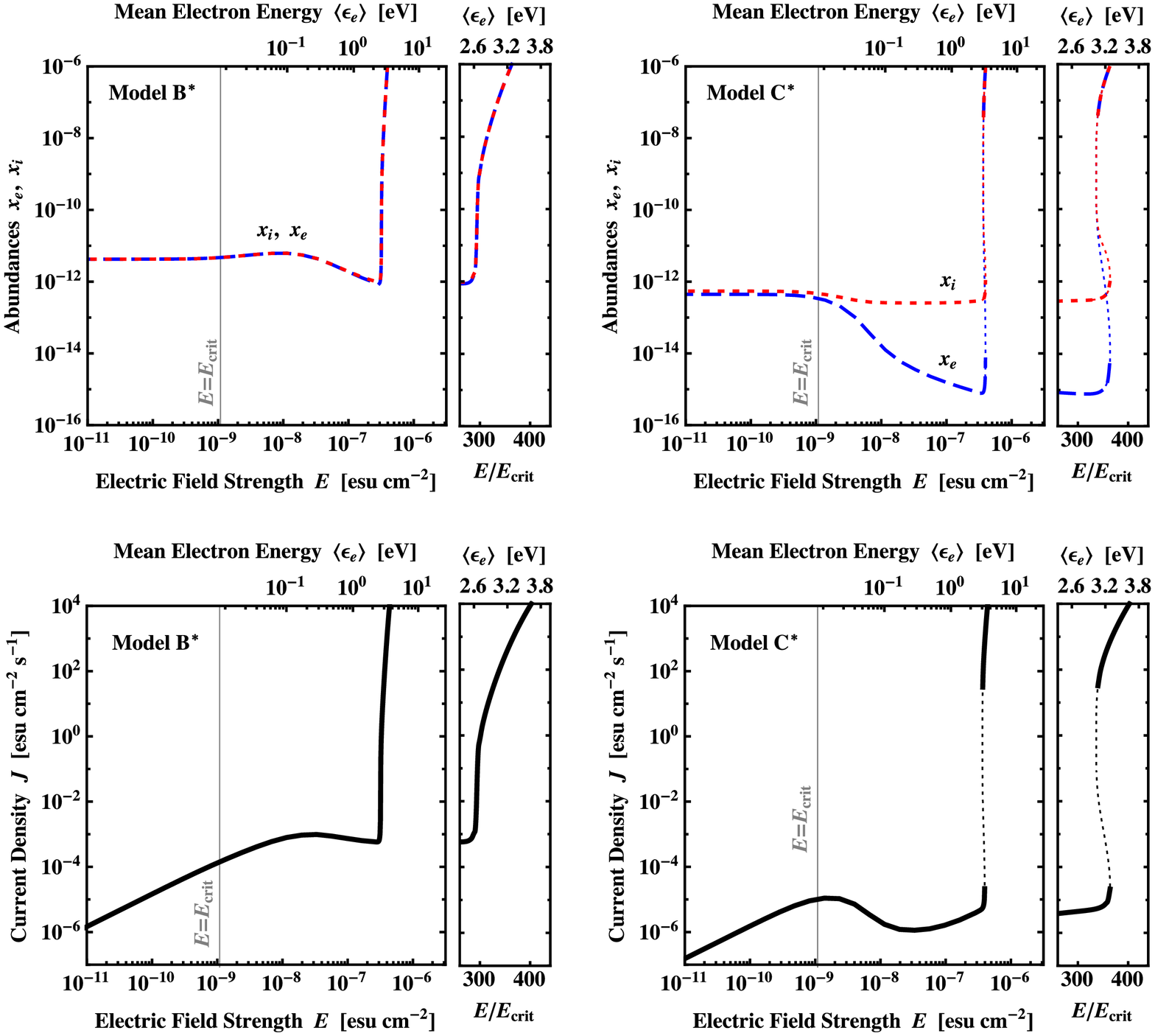}
\caption{Charge abundances (upper panels) and current density (lower panels) 
as a function of the field strength $E$ for models \Bstar~(left panels) and \Cstar~(right panels).
The narrow panels zoom in on the discharge current at $E \approx 260$--$440\Ecrit$.
The dotted curve segments indicate the unstable middle solutions.
}
\label{fig:BC2}
\end{figure*}

Figure~\ref{fig:BC2} shows the plasma abundances $x_i$ and $x_e$ and current density $J$ 
as a function of $E$ for models \Bstar~and \Cstar. 
As expected by previous studies \citep{IS05,MOI12},
impact ionization dramatically changes the ionization state at large $E$.
In both models, we observe an ``electric discharge,'' an abrupt increase in the plasma abundance,
when the mean electron energy $\brac{\eps_e}$ reaches $\approx 3~{\rm eV}$.
This is due to the rapid increase in the impact ionization rate coefficient $K_*$ 
around that electron energy (see Figure~\ref{fig:Kstar}). 
It is interesting to see that the discharge current appears much earlier than
$\brac{\eps_e}$ reaches the ionization potential ${\rm IP} = 15.4~{\rm eV}$.
This means that the high-energy tail of the energy distribution function is responsible for this ionization.
At lower $E$, impact ionization has no effect on the ionization state, {so the curves' left ends 
in models \Bstar~and \Cstar~are identical to those in models B and C, respectively.}

However, the nature of the discharge current is much more complex than 
assumed in the previous studies. 
In model \Cstar, we find that Equation~\eqref{eq:master} has 
{\it three} equilibrium solutions for a single value of $E$ 
when the mean electron energy falls within 
the narrow range $3.0~{\rm eV} \la \brac{\eps_e} \la 3.3~{\rm eV}$.
The triple solution forms an S-curve in the $J$--$E$ space
as shown in the small panel of Figure~\ref{fig:BC2}.
By contrast, in model \Bstar, the discharge current is single-valued for all $E$. 
In the following subsections, we analyze the structure of the equilibrium solutions in more detail. 

\subsection{Classification of Equilibria}\label{sec:avalanche}
 
\begin{figure*}[t]
\epsscale{1.0}
\plotone{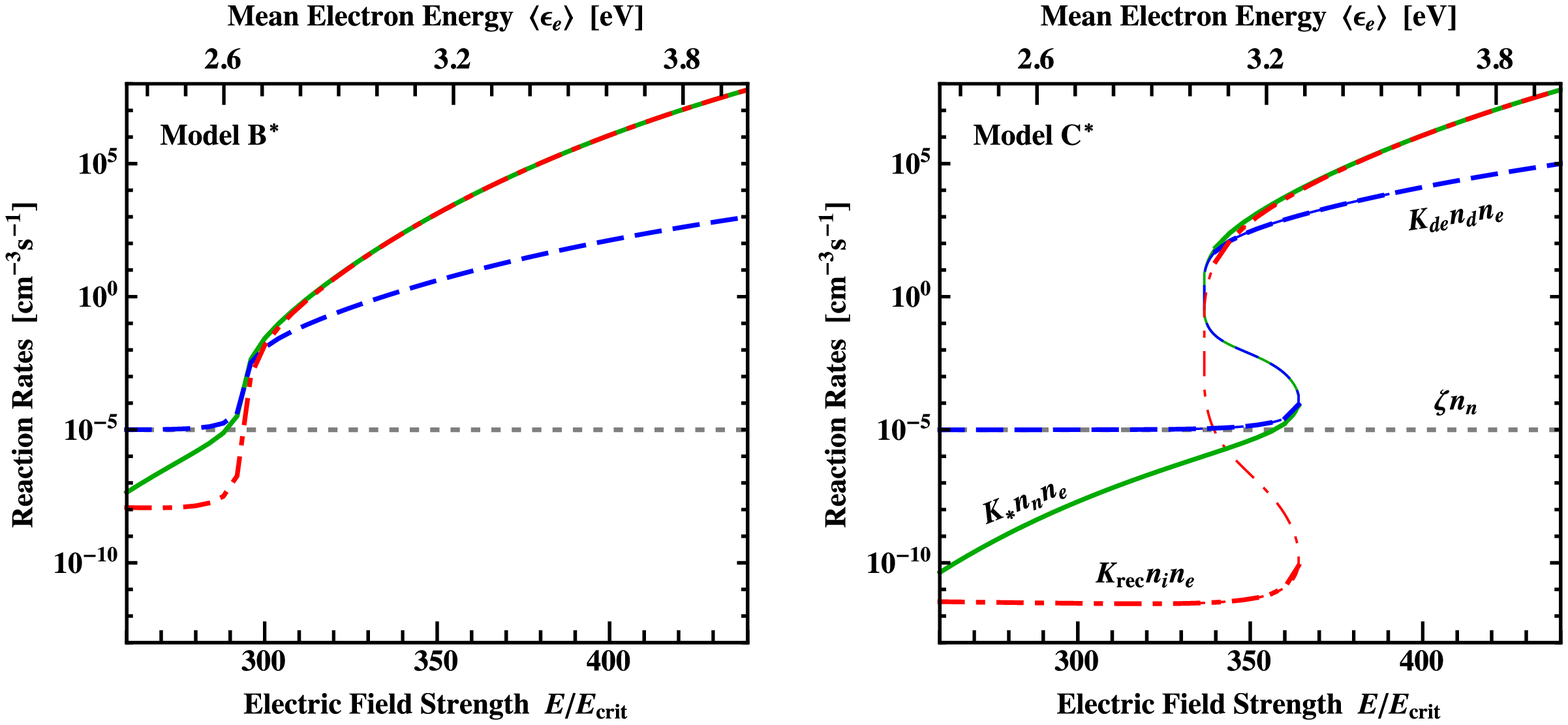}
\caption{Reaction rates of external ionization ($\zeta n_n$; dotted curves), 
impact ionization ($K_* n_n n_e$; solid curves), 
gas-phase recombination ($K_{\rm rec}n_in_e$; dot-dashed curves), 
and electron sticking onto grains ($K_{de} n_d n_e$; dashed curves) 
for models \Bstar ({left} panel) and \Cstar ({right} panel) as a function of $E$.
In model \Cstar, $J$ is triple-valued at  $340 \la E/\Ecrit \la 360$.
The unstable middle solution is indicated by the thin curves.
}
\label{fig:rates}
\end{figure*}
First let us see how the reaction balance is changed by impact ionization. 
In Figure~\ref{fig:rates}, we plot the rates (per unit volume) of external ionization ($\zeta n_n$), 
impact ionization ($K_* n_n n_e$), gas-phase recombination ($K_{\rm rec}n_in_e$), 
and electron sticking onto grains ($K_{de} n_d n_e$) as a function of $E$. 
In the limit of low $E$, external ionization balances with plasma sticking onto grains
(i.e., $\zeta n_n \approx K_{de} n_d n_e$), 
and impact ionization is negligible as well as gas-phase recombination. 
In the opposite limit, impact ionization balances with gas-phase recombination
(i.e., $K_* n_nn_e \approx K_{de} n_d n_e$), and external ionization and plasma sticking are subdominant 
(in general, gas-phase recombination becomes more and more important 
as the ionization degree is increased).
In the following, we will refer to the former ionization state as the low state, 
and to the latter as the high state.
In model \Bstar, the low state is smoothly connected to the high state at $E \approx 300\Ecrit$. 

In model \Cstar, we can see the third type of ionization state. 
Of the three equilibrium solutions appearing at $340 \la E/\Ecrit \la 360$, 
the top and bottom solutions are merely an extension of the low and high states, respectively,
but the middle solution is characterized by 
the balance between impact ionization and plasma capture by dust grains 
(i.e.,  $K_* n_nn_e \approx K_{\rm rec}n_in_e$). 
We will call this the middle state.

\subsection{Emergence of Multiple Equilibria}
\begin{figure}[t]
\epsscale{1.1}
\plotone{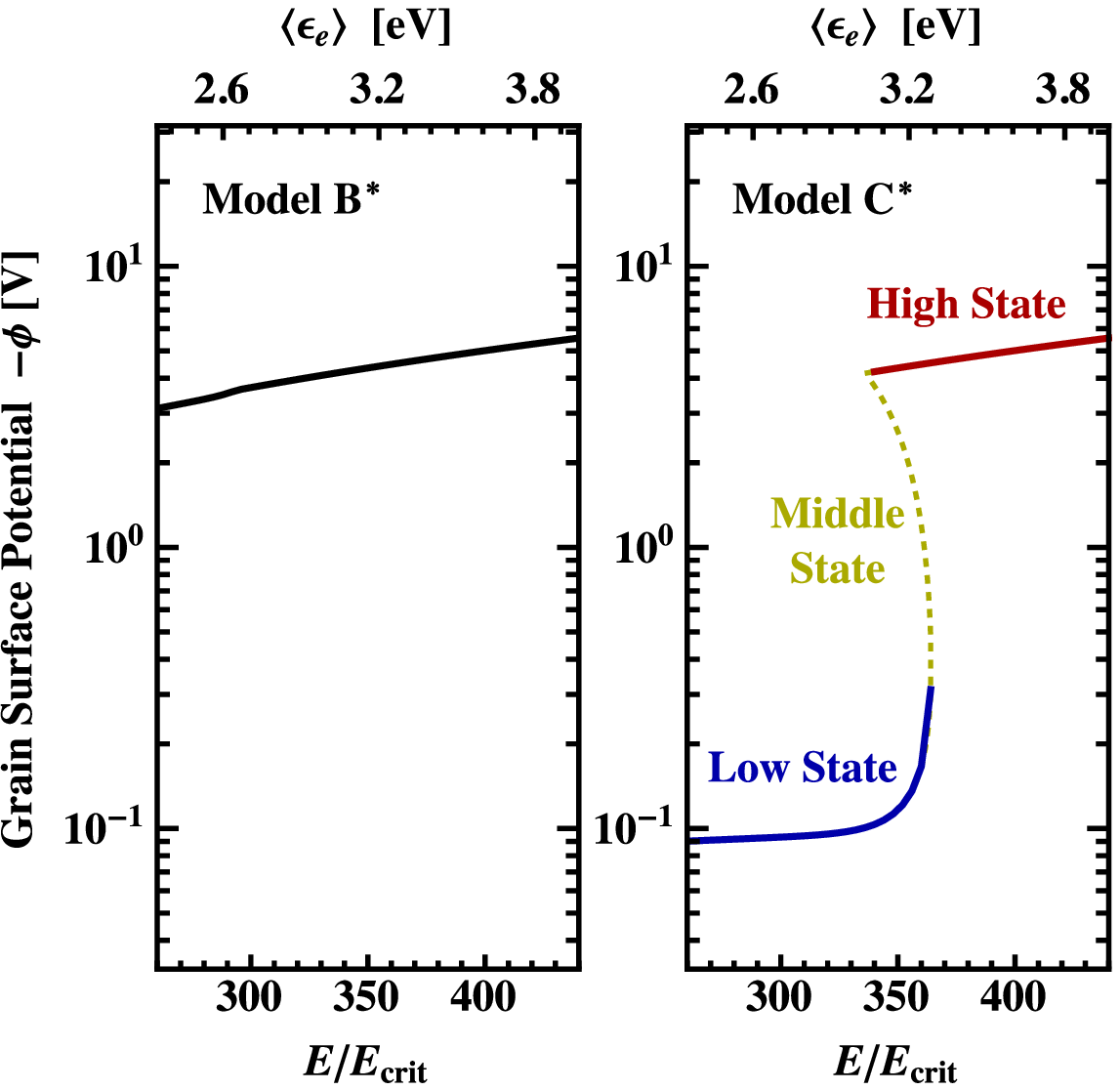}
\caption{Grain surface potential $\phi$ in equilibrium (Equation~\eqref{eq:master})
for models \Bstar~(left panel) and \Cstar~(right panel) as a function of $E$. 
The dotted line indicates the unstable middle solution.
}
\label{fig:phi2}
\end{figure}
It is easy to explain why no multiple equilibrium solution appears in model \Bstar.
We first note that the grain surface potential $\phi$ in equilibrium 
must satisfy the relation 
\beq
K_{di}(E,\phi)n_i = K_{de}(E,\phi)n_e
\label{eq:phi_equil}
\eeq
(see Equation~\eqref{eq:evol_Z}). 
In model \Bstar, gas-phase electrons are so abundant that $n_i \approx n_e$ 
(or equivalently, ${\cal P} \ll 1$) even before the onset of impact ionization.  
For fixed $E$, $K_{di}$ is a decreasing function of $\phi$, 
while $K_{de}$ is an increasing function of $\phi$. 
Therefore, in the case of $n_i \approx n_e$, 
Equation~\eqref{eq:phi_equil} {\it uniquely} specifies $\phi$, 
and in turn $n_e$ ($\approx n_i$), for each value of $E$.
In fact, if we look at $\phi$ of model \Bstar, there is no appreciable change 
in $\phi$ before and after the onset of impact ionization (see the left panel of Figure~\ref{fig:phi2}).   
This is the reason why no multiple solution emerges in model \Bstar.
By contrast, in model \Cstar, the condition $n_i \approx n_e$ is violated 
(or equivalently, ${\cal P} \gg 1$) before the onset of impact ionization. 
In this case, Equation~\eqref{eq:phi_equil} can allow more than one value for $\phi$, 
because the ratio $n_e/n_i$ can vary with $\phi$ 
and hence the monotonicity of Equation~\eqref{eq:phi_equil} in $\phi$ is not ensured.  
This argument suggests that multiple solutions emerge only when the condition ${\cal P} > 1$ 
is satisfied before the onset of electric discharge.

\subsubsection{Instability of the Intermediate (Middle) State}\label{sec:multiple}
\begin{figure}[t]
\epsscale{1.1}
\plotone{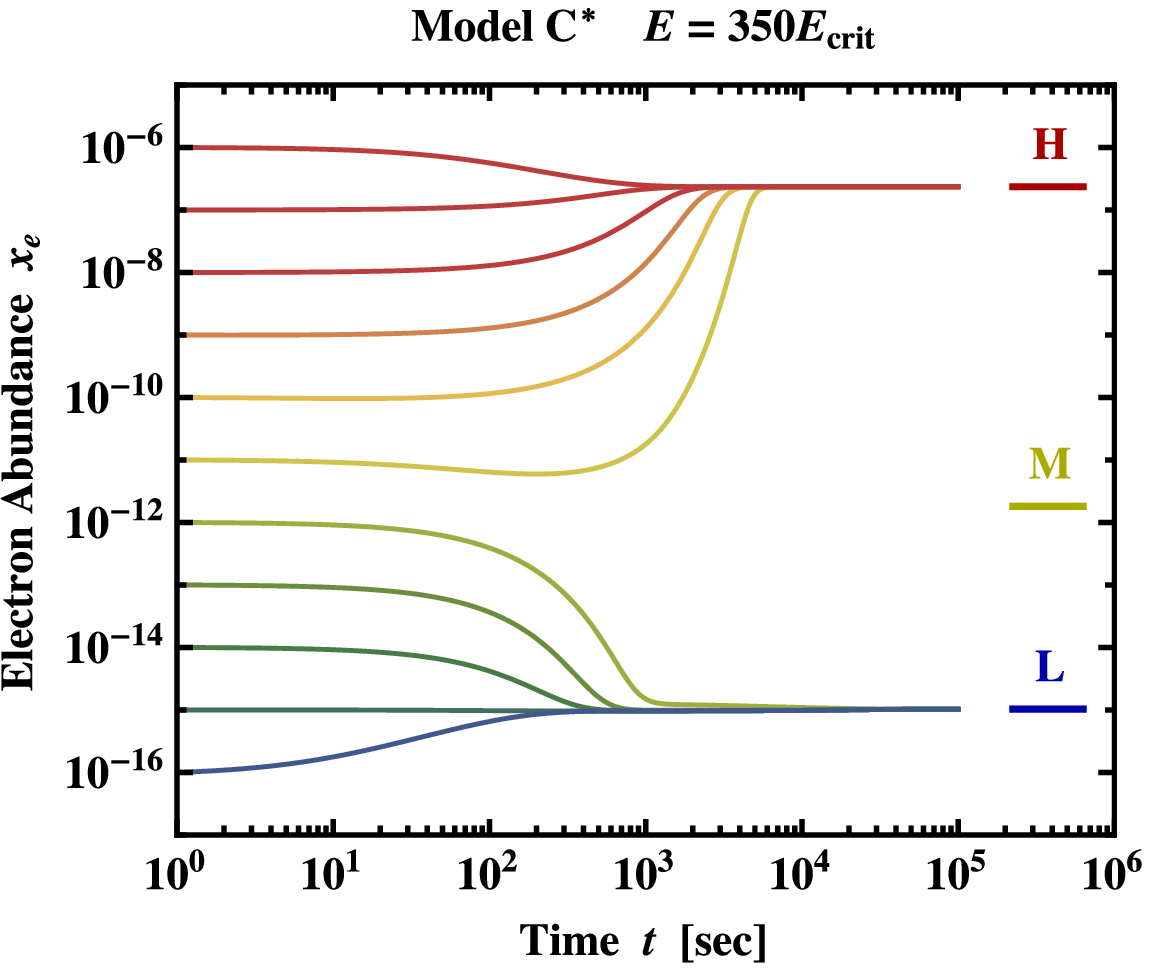}
\caption{Stability check of the triple equilibrium solution in model \Bstar at $E = 350\Ecrit$. The thin curves show the time evolution of the electron abundance $x_e$ 
obtained by integrating the rate equations (Equations~\eqref{eq:evol_ni} and \eqref{eq:evol_ne}) with various initial abundances.
The thick line segments on the right show the three equilibrium solutions
(`L': low state; `M': middle state; `H': high state).
No time-dependent solution approaches the middle state, indicating that the middle state is unstable.
}
\label{fig:xt}
\end{figure}

Emergence of a triple equilibrium or an S-shaped equilibrium curve
can be seen in many physical systems.
In most cases, two extreme equilibria are stable, while the middle equilibrium is unstable against perturbation.
We find that this is also the case for our multiple solutions.
We numerically solved the time-dependent rate equations 
(Equations~\eqref{eq:evol_ni}--\eqref{eq:evol_Z}) for various initial conditions
and looked at which of the equilibria is reached at late times. 
An example of such tests is shown in Figure~\ref{fig:xt}. 
Here we plot the time evolution of the electron abundance $x_e$ for different initial conditions 
in the case of model \Cstar~with $E = 350\Ecrit$.
The low, middle, and high equilibrium states correspond to $x_e \approx 1\times 10^{-15}$, 
$2\times 10^{-12}$, and $2\times 10^{-7}$, respectively. 
As we can see, all the time-dependent solutions converge toward either the high or low state, 
while the middle state is never reached even if the initial state is very close to it.
For completeness, in Appendix~\ref{sec:stability}, we perform a linear stability analysis 
of the middle state by using simplified rate equations, and show that the middle solution is indeed unstable.
From these facts, we conclude that the middle equilibrium of the discharge current is never realized in real systems.

\section{Implications for MHD in Protoplanetary Disks}\label{sec:discussion}
In this section, we discuss important implications of our model calculations 
for the MHD of protoplanetary disks.

\subsection{Negative Differential Resistance and its Instability}\label{sec:NDR}
Negative differential resistance refers to the property of 
some electric circuits (e.g., Gunn diodes) that 
an increase in the applied voltage causes a decrease in the electric current.
Interestingly, some of our model calculations yield 
$J$--$E$ relations that have negative differential resistance (i.e., $dJ/dE<0$) in some range of $E$.
For example, in model C, we see that $J$ decreases by an order of magnitude 
when going from $E \approx 3\Ecrit$ to $E \approx 30\Ecrit$. 
As we will discuss below, negative differential resistance has many important implications 
for the evolution of electric fields and for the MHD of protoplanetary disks.

The most important consequence of negative differential resistance is that 
the displacement current neglected in Amp\`{e}re's law ceases to be negligible. 
To see this, let us consider the Maxwell--Amp\`{e}re equation
\beq
\frac{\pd {\bm E}}{\pd t} = c\nabla\times {\bm B} - 4\pi {\bm J}
\label{eq:MaxwellAmpere}
\eeq
with 
${\bm J} = J(E) \hat{\bm E}$, 
where $J$ is a nonlinear function of the electric field strength $E = |{\bm E}|$.
Note again that all the quantities are defined in the comoving frame of neutrals.
We assume that the background electric field ${\bm E}_0$ 
is imposed by external sources and is approximately steady over a long timescale. 
Then, the background magnetic field ${\bm B}_0$ is related to ${\bm E}_0$ 
by the classical Amp\`{e}re's law
\beq
c\nabla\times{\bm B}_0 = 4\pi {\bm J}({\bm E}_0).
\label{eq:Ampere}
\eeq 
We examine the stability of this relation by considering a perturbation 
${\bm E}(t) = {\bm E}_0 +  {\bm E}_1(t)$, 
where $|{\bm E}_1| \ll |{\bm E}_0|$.
For simplicity, we drop the perturbation of the $c \nabla \times {\bm B}$ term 
by assuming that the wavelength of the perturbed field is sufficiently long.\footnote{It can be shown, by using Equation~\eqref{eq:MaxwellAmpere} and Faraday's law, that the assumption made here is valid if the wavelength of the perturbed field is much longer than $c/|\sigma_{\rm diff}|$, where $\sigma_{\rm diff}$
is the differential conductivity defined by Equation~\eqref{eq:sigma_diff}. }  
Substituting this into Equation~\eqref{eq:MaxwellAmpere} 
we obtain 
\beq
\frac{\pd {\bm E}_1}{\pd t} = - 4\pi \sigma_{\rm diff}{\bm E}_1,
\label{eq:dE1dt}
\eeq
where 
\beq
\sigma_{\rm diff} \equiv \frac{dJ}{dE} (|{\bm E}_0|)
\label{eq:sigma_diff}
\eeq 
is the differential conductivity evaluated at $E = |{\bm E}_0|$.
To order of magnitude, $|\sigma_{\rm diff}|^{-1} \sim E/J$, 
which is $\sim 10^{-4}$--$10^{-2}~{\rm s}$ 
for models \Bstar~and \Cstar~in the region of negative $\sigma_{\rm diff}$.
If $\sigma_{\rm diff}$ is positive, as is the case for the conventional linear Ohm's law, 
the perturbation {in $E$} decays on a timescale of $(4\pi \sigma_{\rm diff})^{-1}$ 
(which is known as the Faraday time).
However, if $\sigma_{\rm diff}$ is negative, the perturbation grows exponentially with time,
meaning that the Amp\`{e}re's law (Equation~\eqref{eq:Ampere}) is unstable. 

\begin{figure}[t]
\epsscale{1.1}
\plotone{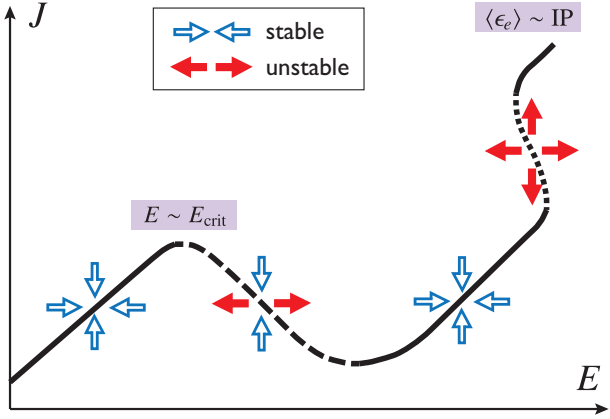}
\caption{Summary of the stability of the nonlinear Ohm's law.  
The black curve schematically shows the equilibrium $J$--$E$ relation for model \Cstar.
The dashed line indicates negative differential resistance, and 
the dotted line the middle branch of the discharge current. 
The open (filled) arrows indicate that the equilibrium  
is stable (unstable) against perturbations in the direction of the arrows.
Note that whether the negative differential resistance and triple-valued discharge appear
depends on the model parameters.
}
\label{fig:instability}
\end{figure}
Figure~\ref{fig:instability} summarizes the instabilities of the nonlinear Ohm's law identified in this study. 
As already seen in Section~\ref{sec:NDR}, the middle branch of the discharge current has, 
whenever present, an unstable ionization balance. 
In Figure~\ref{fig:instability}, the unstable middle branch is indicated by the dotted line, 
and the instability by the vertically diverging arrows.
In addition to this, we have found here that a current with a negative $dJ/dE$ 
is unstable to perturbations in $E$. A negative $dJ/dE$ can appear at $E \ga \Ecrit$ 
and at the discharge current as shown by the dashed and dotted line segments in Figure~\ref{fig:instability}, 
respectively. The associated instability is represented by the horizontally diverging arrows.

The instability of negative differential resistance has a significant impact on the MHD of the system.
The fundamental assumption of the standard non-relativistic MHD is that 
the displacement current is negligible (i.e., Amp\`{e}re's law approximately holds) 
on a dynamical timescale.
As shown above, this assumption is always valid for linear Ohm's laws, but not for 
nonlinear Ohm's laws that exhibit negative differential resistance over some range of $E$. 
In the latter case, one must in principle treat the dynamics of the system by
using the fully time-independent Maxwell-Amp\'{e}re's law (Equation~\eqref{eq:MaxwellAmpere}) 
instead of the quasi-steady Amp\'{e}re's law.
Of course, such a task is computationally challenging 
since one then needs to treat unwanted electromagnetic modes that appear at the same time.  

It should be noted, however, that the linear analysis presented above 
does not directly apply to real protoplanetary disks.
In a typical disk environment, the predicted timescale of the instability ($\ll 1~{\rm s}$) 
is much shorter than the relaxation timescales of the velocity distributions and  
charge reactions of the plasmas. 
This invalidates the use of the relation $J = J(E)$,  
which assumes that both $n_\alpha$ and $\brac{{\bm v}_{||\alpha}}$ ($\alpha = i, e$) 
instantaneously reach an equilibrium for given $E$.
To go further, we need to 
consider non-equilibrium evolution of $J$ instead of using the quasi-equilibrium relation $J = J(E)$.  
Such a task is beyond the scope of this paper, but will be addressed in our future work.

Nevertheless, we may argue that inclusion of the displacement current 
is essential to treat the dynamics of the system with negative differential resistance.
When we neglect the displacement current, we lose the ability to 
treat the electric field ${\bm E}$ as an independent dynamical quantity 
since the displacement current is responsible for the time evolution of ${\bm E}$.
This is not an issue when ${\bm J}$ is monotonic in ${\bm E}$,
because ${\bm E}$ is then uniquely determined as a function of ${\bm J}$. 
However, in the presence of negative differential resistance, 
${\bm E}$ becomes multivalued as a function of ${\bm J}$,
and therefore cannot be determined by instantaneous relations. 
In this case, the state of the system depends on the {\it history} of ${\bm E}$, 
and this can only be determined by Maxwell-Amp\'{e}re's equation with the $\pd {\bm E}/\pd t$ term.
In a forthcoming paper, we will show that the displacement current 
naturally solves this issue by allowing hysteresis for the relation between ${\bm J}$ and ${\bm E}$.

\subsection{Implications for MRI Turbulence}\label{sec:PPD}
One important finding of this study is that plasma heating reduces 
the electric conductivity $J/E$ before the onset of impact ionization. 
This was not considered in our previous studies \citep{IS05,MOI12}, 
which only assumed that plasma heating enhances the conductivity via impact ionization.
The reduction of the conductivity might lead to self-regulation 
of the magnetohydrodynamic motion of the gas: 
coupling between the moving gas and a magnetic field generates 
an electric field in the comoving frame, but this causes a reduction of the conductivity
and hence the coupling between the gas and magnetic field.

Such an effect is of particular importance to MRI turbulence in protoplanetary disks 
as it could limit or even determine the saturation level of the turbulence. 
However, in order to prove this, we would have to perform a resistive MHD simulation including the effect 
of plasma heating on the resistivity, which is clearly beyond the scope of this paper. 
Below, we shall only speculate, by using two illustrative examples, 
how MRI turbulence will develop in a protoplanetary disk under the effect of plasma heating.

As mentioned in Section~\ref{sec:estimate}, 
fully saturated MRI turbulence is characterized 
by the universal average current density $J_{\rm MRI}$ (Equation~\eqref{eq:JMRI}).
Therefore, MRI tends to grow until the current density $J$ reaches this saturation value.
However, MRI does not grow but decays when the Elsasser number 
$\Lambda$ (Equation~\eqref{eq:Lambda}) is less than the critical value $\Lambda_{\rm crit} \sim 0.1$--1. 
Taken together, we may assume that MRI grows until either $J$ reaches $J_{\rm MRI}$
or $\Lambda$ falls below $\Lambda_{\rm crit}$. 

Let us map these criteria onto the $J$--$E$ diagrams presented in this study.
From Equations~\eqref{eq:JMRI}, $J_{\rm MRI}$ can be evaluated as 
\beq
J_{\rm MRI} \approx 2\times 10^{-3} \pfrac{f_{\rm sat}}{10}\pfrac{n_n}{10^{12}~{\rm cm^{-3}}}^{1/2}\pfrac{30~{\rm yr}}{t_K}~{\rm esu~cm^{-2}~s^{-1}},
\label{eq:MRI_PPD}
\eeq
where $t_{\rm K}$ is the local orbital period.   
The MRI stability criterion $\Lambda < \Lambda_{\rm crit}$ can be rewritten as 
the condition for the conductivity $J/E$, 
\beqn
\frac{J}{E} &\la& \frac{\beta_z\Lambda_{\rm crit}\Omega}{8\pi}\pfrac{c}{c_s}^2
\nonumber \\
&\approx& 2\times 10^{4}\Lambda_{\rm crit}\pfrac{\beta_z}{100}\pfrac{100~{\rm K}}{T}
\pfrac{30~{\rm yr}}{t_K}~{\rm s^{-1}}.
\label{eq:DZ}
\eeqn
The orbital period of $t_{\rm K} = 30~{\rm yr}$ corresponds to 
the distance of $\approx 10~{\rm AU}$ from a solar-mass star.
Below we assume $f_{\rm sat} = 10$ and $\Lambda_{\rm crit} = 1$.

\begin{figure*}[t]
\epsscale{1.1}
\plotone{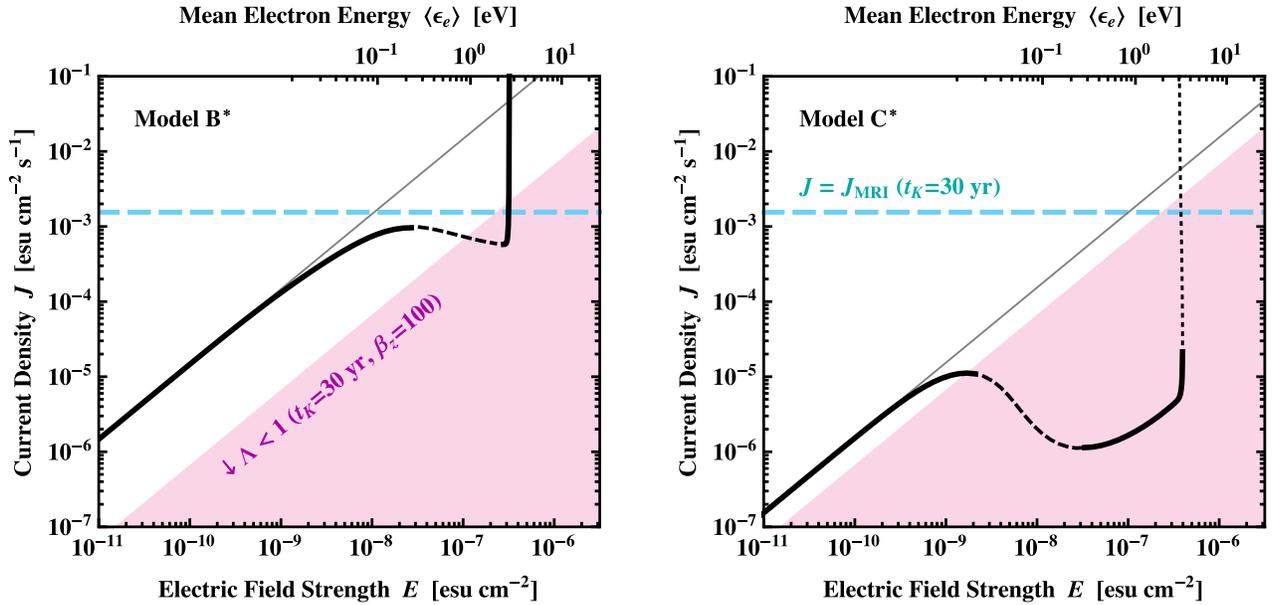}
\caption{$J$--$E$ diagrams (black curves) for models \Bstar~(left panel) and \Cstar~(right panel), 
mapped with the MRI current $J_{\rm MRI}$ (Equation~\eqref{eq:MRI_PPD}; long-dashed lines) 
and MRI stability criterion $\Lambda<1$ (Equation~\eqref{eq:DZ}; shaded regions) for $t_K = 30~{\rm yr}$ 
and $\beta_z = 100$.
The gray lines show the conventional linear Ohm's laws.
The short-dashed and dotted portions of the black curves indicate unstable branches 
(see Figure~\ref{fig:instability}).
}
\label{fig:PPD}
\end{figure*}
In Figure~\ref{fig:PPD}, we plot the equilibrium $J$--$E$ relations for models \Bstar~and 
\Cstar~together with the saturation criterion $J=J_{\rm MRI}$ and stability criterion $\Lambda<1$.
Here we assume $t_K = 30~{\rm yr}$, so that the parameter set ($n_n$, $T$, $t_K$) 
approximately corresponds to the orbital distance of $10~{\rm AU}$ in the minimum-mass solar nebula.
The value of $\beta_z$ is taken to be $100$, which corresponds to 
strong turbulence induced by a large net poloidal flux \citep[see, e.g.,][]{HGB95}.
For comparison, the conventional linear Ohm's law is also plotted.
In both models, the instability criterion $\Lambda > 1$ is satisfied 
at $E < \Ecrit~(\approx 10^{-9}~{\rm esu~cm^{-2}})$, 
so MRI is active at least in its early growth stages. 
%
{However, at $E > \Ecrit$,  $J$} starts to decrease and 
cross the $\Lambda = 1$ line before it reaches $J_{\rm MRI}$. 
Since MRI grows at $\Lambda > 1$ and decays at $\Lambda < 1$, 
we expect that the turbulence will saturate on the $\Lambda = 1$ line. 
{In particular,} the saturation level {in model \Bstar} is expected to be 
much lower than that of fully developed MRI turbulence
because the value of $J$ at $\Lambda = 1$ is two orders of magnitude smaller than $J_{\rm MRI}$
(i.e., the value of $f_{\rm sat}$ introduced in Section~\ref{sec:estimate} is as small as $0.1$).
Of course, a quantitative estimate of the saturation level requires MHD simulations.

\section{Summary and Conclusions}\label{sec:summary}
MRI generates strong electric fields, and such fields can significantly heat up 
plasmas in weakly ionized protoplanetary disks.
To study how this affects the ionization state and MHD of the disks,
we have formulated a charge reaction model that takes into account plasma 
heating and impact ionization by heated electrons
as well as plasma accretion onto dust grains. 
The output of our model is the electric current density $J$ 
as a function of the electric field strength $E$ as measure in the comoving frame of the neutral gas.
Because the plasma heating changes the ionization degree of the gas,
the resulting Ohm's law is {\it nonlinear} in $E$. 

We have presented some model calculations to illustrate
the effects of plasma heating on the ionization balance of a dusty gas. 
The key findings are summarized as follows.

\begin{enumerate}
\item 
When impact ionization is negligible, 
the ionization states are characterized by 
(1) which of gas-phase recombination and plasma sticking to grains  
(``solid-phase recombination'') dominates the reaction balance, 
and by (2) which of gas-phase electrons and charged grains 
are the dominant negative charge carriers. 
These two conditions are quantified by the dimensionless 
{grain} recombination parameter ${\cal S}$ (Equation~\eqref{eq:calS}) 
and Havnes parameter ${\cal P}$ (Equation~\eqref{eq:calP}), respectively. 
For both conditions, the presence of dust becomes more and more important 
(both ${\cal S}$ and ${\cal P}$ increase) as the number of small dust grains increases, 
the external ionization rate $\zeta$ decreases, and/or the electric field strength $E$ 
increases (Figure~\ref{fig:SP}). 
The field strength is relevant 
because electrons hit and stick to dust grains more and more frequently
as they are heated up. 

\item
When plasma accretion by dust grains dominates over gas-phase recombination 
(${\cal S}>1$), the electron abundance decreases with increasing $E$ 
(Section~\ref{sec:ionization}, Figure~\ref{fig:x}) because 
of the electron--grain collisions facilitated by the electron heating.
The current density $J$ also decreases 
until the electron current is taken over by the ion current (Section~\ref{sec:current} and Figure~\ref{fig:J}). 
In particular, $J$ rapidly decreases when charged grains replace free electrons as  
the dominant negative charge carriers of the system (i.e., when ${\cal P}$ crosses unity). 
These results have very important implications for the MHD of the system.
First, the decrease of the electron abundance implies that MRI turbulence can be self-regulating: 
as MRI grows, the magnetic resistivity increases, which prevents further growth of MRI (Section~\ref{sec:PPD}).  
Furthermore, the N-shaped $J$--$E$ curve 
violates the fundamental assumption of the standard non-relativistic MHD
that a single value of $J$ corresponds to a single value of the comoving field strength $E$. 
In fact, our simple linear analysis suggests that 
the negative differential resistance $(dJ/dE < 0)$ should destabilize the electric field via 
the displacement current, which implies that the dynamical evolution of the system should 
depend on the history of the electric field (Section~\ref{sec:NDR}). 

\item Impact ionization by hot electrons sets in 
when the mean electron energy exceeds a few eV.
This results in an abrupt increase in the electric current 
as previously investigated by \citet{IS05} and \citet{MOI12} (Section~\ref{sec:discharge}).
We find that this discharge current is triple-valued as a function of $E$
(i.e., the $J$--$E$ curve is S-shaped) 
when charged dust grains dominate the charge neutrality (${\cal P} > 1$) at low $E$.
Furthermore, the middle branch of the S-shape current is found to 
be unstable to perturbations to the ionization balance.
Therefore, the MHD near the discharge current could be more complex 
than self-sustained turbulence as envisaged by \citet{IS05} and \citet{MOI12}.
\end{enumerate}

Plasma heating could also have a significant influence on the collisional growth of dust grains. 
Since grains in a plasma have a nonzero (and negative) mean charge, 
their collisional cross section is on average smaller than their geometric cross section.
This ``charge barrier'' can slow down the growth of small dust grains even 
in weakly ionized protoplanetary disks \citep{O09,OTTS11b,MLH12}.
The maximum value of the grain negative surface potential $-\phi$ 
is given by the energy balance $-e\phi \sim \brac{\eps_e}$ (see also our Figure~\ref{fig:phi}). 
The maximum negative potential is $\sim -10~{\rm mV}$ in a cool gas of $T \sim 100~{\rm K}$, 
but can exceed $\sim -1~{\rm V}$ when the plasma is heated by a strong electric field. 
Because the Coulomb repulsion energy between two grains is $\propto \phi^2$, 
the heating can lead to a $\sim 10^4$-fold enhancement of the repulsion energy. 
Therefore, in a disk region where plasma heating is effective, 
the growth of dust grains could be more strongly suppressed than previously thought.  

A major limitation of this study is that
the velocity distribution functions adopted {here} neglect 
the effects of magnetic fields on the kinetics of the plasmas.
In terms of non-ideal MHD, we have only considered ohmic diffusivity 
and neglected ambipolar diffusion and Hall drift.
Such a treatment is only valid in dense gases  
where the frequency of plasma--neutral collisions is much higher than the gyration frequency of the plasmas \citep[e.g.,][]{NU86a,W99}. 
If the neutral drag is weak, plasma particles undergo gyromotion, 
which prevents plasma heating when the electric field is perpendicular to the magnetic field \citep{GZS80}.
This effect is non-negligible over a wide region of protoplanetary disks 
where Hall drift or ambipolar diffusion dominates over ohmic diffusion \citep{BT01,W07,B11a}. 
Our future {modeling} will take into account this effect. 

\acknowledgments
We thank Takayuki Muranushi, Xuening Bai, Takeru Suzuki, Shigenobu Hirose, 
Hidekazu Tanaka, Naoki Watanabe, Shota Nunomura, Takayuki Muto, and Shoji Mori
for useful comments and inspiring discussions.
We also thank the referee, Neal Turner, for his prompt and detailed report 
that greatly improved our presentation.
This work is supported by Grant-in-Aid for Research Activity Start-up (\#25887023) from JSPS, 
and by Grants-in-Aid for Scientific Research (\#23103005) from MEXT. 

\appendix

\section{Kinetics of Weakly Ionized Plasmas under an Electric Field}
\label{sec:kinetics}
In this section, we briefly review the kinetics of weakly ionized plasmas under an applied electric field. 
A more comprehensive review can be found in \citet{W53} and in \citet{GZS80}.

We consider the motion of charged particles in a neutral gas in the presence of 
an applied $E$-field.
We assume that the number density $n_\alpha$ of the charged particles 
are so low that collisions between the charged particles are rare.
In this case, the charged particles gain and lose their momentum $m_\alpha {\bm v}_\alpha$
and kinetic energy $\eps_\alpha$ through collision with neutrals and acceleration by
the electric fields.
The equations of momentum and energy balance are given by \citep[e.g.,][]{H39}
\beq
m_\alpha\dbrac{\frac{\Delta{\bm v}_\alpha}{\Delta t_\alpha} } + q_\alpha{\bm E} = 0,
\label{eq:balance_p}
\eeq
\beq
\dbrac{ \frac{\Delta\eps_\alpha}{\Delta t_\alpha} } + q_\alpha{\bm E}\cdot\brac{{\bm v}_{\alpha||}} = 0.
\label{eq:balance_e}
\eeq
Here, $\Delta{\bm v}_\alpha$ and $\Delta\eps_\alpha$ are the change in 
the changes in ${\bm v}_\alpha$ and $\eps_\alpha$ upon individual collisions 
averaged over the scattering angle, respectively, 
$\Delta t_\alpha$ is the mean free time of the charged particles, 
and the double brackets $\dbrac{\cdots}$ stand for the average 
over the charged and neutral particle velocities.
Note that $\Delta t_\alpha$ is generally a function of the relative speed 
$|{\bm v}_\alpha - {\bm v}_n|$.
If the collision with neutrals is elastic and isotropic, 
$\Delta{\bm v}_\alpha$ and $\Delta\eps_\alpha$ can be written as \citep[e.g.,][]{GZS80}
\beq
\Delta{\bm v}_\alpha
= - \lambda_{\alpha n}({\bm v}_\alpha- {\bm v}_n),
\label{eq:Deltavalpha_C}
\eeq
\beq
\Delta\eps_\alpha
= -\kappa_{\alpha n}\left[\eps_\alpha -\eps_n 
-\frac{1}{2}(m_\alpha-m_n){\bm v}_\alpha\cdot{\bm v}_n \right],
\label{eq:Deltaealpha_C}
\eeq
where
\beq
\lambda_{\alpha n} \equiv \frac{m_n}{m_\alpha+m_n},
\eeq
\beq
\kappa_{\alpha n} \equiv \frac{2m_\alpha m_n}{(m_\alpha+m_n)^2}
\eeq
 are the momentum and energy transfer efficiencies for the elastic collisions, respectively. 
Note that $\lambda_{\alpha n} \ll 1$ for heavy charged particles ($m_\alpha \gg m_n$), 
while $\lambda_{\alpha n} \approx 1$ for intermediate-mass and light charged particles 
($m_\alpha \la m_n$).
By contrast,  $\kappa_{\alpha n} \ll 1$
for both light and heavy charged particles, 
and  $\kappa_{\alpha n} \sim 1$ only for $m_\alpha \sim m_n$.
For electrons ($n_e \ll m_n$), a single collision perfectly isotropizes 
the velocity distribution of the electrons but hardly affects their energy distribution. 
For heavy ions ($m_i \gg  m_n$), a single collision with a neutral 
hardly changes the momentum and energy of the ions.

It is known that the mean free time of ions is approximately constant 
(i.e., independent of $|{\bm v}_\alpha - {\bm v}_n|$) because of the polarization force acting between ions and neutrals \citep[see, e.g.,][]{W53}.
In this case, Equations~\eqref{eq:balance_p} and \eqref{eq:balance_e} 
are closed with respect to $\brac{{\bm v}_{\alpha||}}$ and $\brac{\eps_{\alpha}}$,
and we obtain 
\beq
- m_\alpha\lambda_{\alpha n}  \brac{{\bm v}_{\alpha||}} 
+ q_\alpha {\bm E} \Delta t_\alpha = 0,
\label{eq:balance_p2}
\eeq
\beq
\kappa_{\alpha n}
\left( \frac{3\kB T}{2} - \brac{\eps_\alpha} \right)
+ q_\alpha{\bm E}\cdot\brac{{\bm v}_{\alpha||}}\Delta t_\alpha = 0,
\label{eq:balance_e2}
\eeq
where we have used that $\brac{{\bm v}_n}= 0 $ and $\brac{\eps_n} = 3\kB T/2$.
Solving these equations for $\brac{{\bm v}_{\alpha||}}$ and $\brac{\eps_{\alpha}}$, 
we obtain
\beq
\brac{{\bm v}_{\alpha||}} 
=  \dfrac{q_\alpha{\bm E}\Delta t_\alpha}{m_\alpha\lambda_{\alpha n} } 
= \dfrac{m_\alpha+m_n}{m_\alpha m_n}q_\alpha{\bm E}\Delta t_\alpha,
\label{eq:valpha||}
\eeq
\beqn
\brac{\eps_\alpha}
&=& \dfrac{3}{2}\kB T + \dfrac{(q_\alpha E \Delta t_\alpha)^2}{m_\alpha\kappa_{\alpha n}\lambda_{\alpha n}}
\nonumber \\
&=& \dfrac{3}{2}\kB T + \dfrac{(m_\alpha+m_n)^3}{2(m_\alpha m_n)^2}
(q_\alpha E \Delta t_\alpha)^2.
\label{eq:ealpha}
\eeqn
In Equation~\eqref{eq:ealpha}, the first term comes from the collisional heating with 
neutrals, while the second term from the field heating.
Equation~\eqref{eq:ealpha} implies that the field heating dominates when 
\beq
E >  \frac{m_\alpha m_n \sqrt{3\kB T}}{(m_\alpha+m_n)^{3/2}|q_\alpha|\Delta t_\alpha} 
\equiv E_{{\rm crit},\alpha}.
\label{eq:Ecrit_ion}
\eeq
If we eliminate $q_\alpha E\Delta t_\alpha$ in \eqn{ealpha} by using \eqn{valpha||},
we obtain an interesting relation
\beq
\brac{\eps_\alpha}  = \frac{3}{2}\kB T + \frac{1}{2}m_n \brac{{\bm v}_{\alpha||}}^2
+ \frac{1}{2}m_\alpha \brac{{\bm v}_{\alpha||}}^2.
\label{eq:ealpha2}
\eeq
Since the third term in Equation~\eqref{eq:ealpha2} is the energy associated 
with the mean drift motion, the sum of the first and second terms represents 
the energy of random motion.
It follows that, in the limit of high $E$ (or low $T$), 
the random energy is exactly $m_n/m_\alpha$ times the drift energy,
and the ratio of the root-mean-squared speed to the drift speed approaches 
\beq
\frac{\brac{v_\alpha^2}^{1/2}}{|\brac{{\bm v}_{\alpha||}}|} 
\to \sqrt{1+\frac{m_n}{m_\alpha}}  \quad(E \gg E_{\rm crit,\alpha}).
\label{eq:vratio_appendix}
\eeq

If $\Delta t_\alpha$ depends on $|{\bm v}_\alpha-{\bm v}_n|$,
Equations~\eqref{eq:balance_p} and \eqref{eq:balance_e} 
are not closed with respect to $\brac{{\bm v}_{\alpha||}}$ and $\brac{\eps_{\alpha}}$. 
For example, electrons have a mean free time that is nearly inversely proportional 
to $|{\bm v}_e-{\bm v}_n|$, 
because its momentum-transfer cross section (or equivalently, its mean free path $\ell_e = |{\bm v}_e-{\bm v}_n|\Delta t_e$) 
depends on $|{\bm v}_e-{\bm v}_n|$ very weakly \citep[e.g.,][]{FP62,YSH+08}.
Nevertheless, one can use Equations~\eqref{eq:balance_p2} and \eqref{eq:balance_e2} (and hence Equations~\eqref{eq:valpha||} and \eqref{eq:ealpha}) for electrons
to a good approximation if one approximates the electron mean free time $\Delta t_e$ 
as $\ell_e/\sqrt{\brac{v_e^2}} = \ell_e/\sqrt{2\brac{\eps_e}/m_e}$ \citep{W53}.
The approximate equations can be solved with respect to 
$\brac{{\bm v}_{e||}}$ and $\brac{\eps_e}$; in the limit of high fields, the result is  
\beq
\brac{{\bm v}_{e||}} \approx 
-\pfrac{\kappa_{en}}{2}^{1/4} \sqrt{\frac{eE\ell_e}{m_e}}\hat{{\bm E}}
\approx  - \frac{\sqrt{eE\ell_e}}{(m_em_n)^{1/4}}\hat{{\bm E}},
\label{eq:ve||approx}
\eeq
\beq
\brac{\eps_e} 
\approx \frac{eE\ell_e}{\sqrt{2\kappa_{en}}}
\approx \frac{1}{2}\sqrt{\frac{m_n}{m_e}}eE\ell_e,
\label{eq:ee_approx}
\eeq
where we have used that $\lambda_{en} \approx 1$ and $\kappa_{en} \approx 2m_e/m_n$.
Comparing Equation~\eqref{eq:ee_approx} with the weak-field expression 
$\brac{\eps_e} \approx 3\kB T/2$, 
we find that significant electron heating occurs when 
\beq
E \gg  \frac{\sqrt{\kappa_{en}}\kB T}{e \ell_e} \sim \Ecrit.
\eeq
Equations~\eqref{eq:ve||approx} and \eqref{eq:ee_approx} agree with the asymptotic  
expressions of $\brac{{\bm v}_{e||}}$ and $\brac{\eps_e}$ in the limit of $E \gg \Ecrit$ 
(Equations~\eqref{eq:ve||} and \eqref{eq:ee}) 
within an relative error of only 11\% and 17\%, respectively. 
Equation~\eqref{eq:vratio_appendix} also holds within an error of only 3\%
(see Equation~\eqref{eq:ve||ve}).

\section{Stability Analysis of the Middle Solution}\label{sec:stability}
Here, we prove the instability of the middle solution 
using a standard linear perturbation theory.
The middle solution is characterized by the balance between 
 the ionization by high-energy electrons and the sticking of ions and electrons onto grains.
With this fact in mind, 
we may simplify Equations~\eqref{eq:evol_ni} and \eqref{eq:evol_ne} as
\beq
\frac{dn_i}{dt} = -K_{di}(Z)n_dn_i + K_* n_n n_e,
\label{eq:evol_ni_inter}
\eeq
\beq
\frac{dn_e}{dt} = -K_{de}(Z)n_dn_e + K_* n_n n_e,
\label{eq:evol_ne_inter}
\eeq
respectively.
The equilibrium conditions  $dn_i/dt = 0$ and $dn_e/dt =0 $ give 
\beq
K_{di}n_dn_i = K_*n_nn_e,
\label{eq:equil_ni}
\eeq
\beq
K_{de}n_d = K_*n_n,
\label{eq:equil_ne}
\eeq
respectively. 

To analyze the stability of the equilibrium solution, 
we consider perturbations $\delta n_i$, $\delta n_e$, and $\delta Z$ 
around the equilibrium values  $n_i$, $n_e$, and $Z$.
Up to the first order in the perturbations, 
Equations~\eqref{eq:evol_ni_inter}, \eqref{eq:evol_ne_inter}, and \eqref{eq:neut}
are written as
\beq
\frac{d}{dt}\delta n_i = -K_{di}n_d\delta n_i - K'_{di}n_dn_i\delta Z + K_* n_n \delta n_e,
\label{eq:evol_dni}
\eeq
\beq
\frac{d}{dt}\delta n_e =  - K'_{de}n_dn_e\delta Z,
\label{eq:evol_dne}
\eeq
\beq
\delta n_i - \delta n_e + n_d \delta Z = 0.
\label{eq:neut_d}
\eeq
where $K'_{d\alpha} \equiv  dK_{d\alpha}(Z)/dZ$.
In deriving Equation~\eqref{eq:evol_dne}, we have used Equation~\eqref{eq:equil_ne}
to eliminate $(-K_{de}n_d+K_*n_n)\delta n_e$.
Substituting $\delta n_i, \delta n_e, \delta Z \propto \exp(st)$ 
into Equations~\eqref{eq:evol_dni}--\eqref{eq:neut_d}, we obtain the equation for $s$,
\beq
s^2 + s(K_{di}n_d + K'_{de}n_e  - K'_{di}n_i) + (K_i - K_*)K'_{de}  n_en_d=0.
\label{eq:s}
\eeq

The equilibrium solution is stable against the charge perturbations
if the roots of Equation~\eqref{eq:s} are all negative. 
However, Equation~\eqref{eq:equil_ni} and \eqref{eq:equil_ne} imply that
\beq
\frac{K_i}{K_*} = \frac{n_e}{n_i},
\label{eq:KiKstar}
\eeq
Since the charge neutrality condition with $Z<0$ requires $n_e<n_i$,
Equation~\eqref{eq:KiKstar} suggests $K_i - K_* < 0$. 
Furthermore, $K'_{de}$ is generally positive 
because the collisional cross section between grains and electrons 
increases with increasing $Z$ (or decreasing $-Z$).
Since $K_i - K_* < 0$ and $K'_{de}>0$,  
the third term in Equation~\eqref{eq:s} is negative,
implying that the equation has one positive and one negative root.
The existence of a positive $s$ means that the middle solution is unstable.

\bibliographystyle{apj}
\bibliography{myrefs_141216}

\end{document}